\documentclass{aa} \usepackage{graphicx} \usepackage{epsfig}
\newcommand{\lkha}{LkH$\alpha$~198}
\newcommand{\mic}{${\mathrm \mu}$m}
\newcommand{\Mdot}{\dot{M}}
\newcommand{\Msun}{M_{\odot}}
\newcommand{\Rsun}{R_{\odot}}
\newcommand\strong[1]{\ifmmode\mathbf{#1}\else\textbf{#1}\fi}
\newcommand{\gtrsim}{\stackrel{>}{~}}

\def\Msun{\ensuremath{M_\odot}}
\def\Rsun{\ensuremath{R_\odot}}
\def\Rstar{\ensuremath{R_\star}}
\def\yr{\ensuremath{\mathrm{yr}}}
\def\AU{\ensuremath{\mathrm{AU}}}
\def\Msunperyr{\ensuremath{\Msun\cdot\yr^{-1}}}

\def\sci#1#2{\ensuremath{{#1}\times 10^{#2}}}

\begin{document}
%

   \title{Close binary companions of the HAeBe stars \lkha, 
Elias~1, HK~Ori and V380~Ori 
\thanks{Based on observations performed with the 6~m telescope of the
Special Astrophysical Observatory, Russia, the 2.2~m ESO/MPG 
telescope at La Silla, 
and with the NASA/ESA {\em Hubble Space Telescope}, obtained from the 
data archive at the 
Space Telescope Institute. STScI is operated by the association of 
Universities for 
Research in Astronomy, Inc. under the NASA contract  NAS 5-26555. }}

   \author{K. W. Smith\inst{1}
          \and
	  Y. Y. Balega\inst{2}
	  \and
	  W. J. Duschl\inst{3}
          \and
	  K.-H. Hofmann\inst{1}
	  \and
	  R. Lachaume\inst{1}
	  \and
	  T. Preibisch\inst{1}
	  \and
	  D. Schertl\inst{1}
	 \and 
	  G. Weigelt\inst{1}
          }

   \offprints{K.Smith, \\  
              \email{kester@mpifr-bonn.mpg.de}}

   \institute{Max-Planck-Institut f\"ur Radioastronomie (MPIfR), Auf dem H\"ugel 69, 53121 Bonn, Germany 
	     \and
	     Special Astrophysical Observatory, Nizhnij Arkhyz, Zelenchuk region, Karachai-Cherkesia, 357147, Russia
	     \and
	     Institut f\"ur Theoretische Astrophysik, Tiergartenstrasse 15, 69121 Heidelberg, Germany 
             }

   \date{Received , ; accepted 19.10.2004}

   \abstract{We present diffraction-limited bispectrum speckle
interferometry observations of four well-known Herbig Ae/Be (HAeBe)
stars, \object{\lkha}, \object{Elias~1}, \object{HK~Ori} and
\object{V380~Ori}. For two of these, \lkha\ and Elias~1, we present
the first unambiguous detection of close companions. The plane of the
orbit of the new \lkha\ companion appears to be significantly inclined
to the plane of the circumprimary disk, as inferred from the
orientation of the outflow.  We show that the Elias~1 companion may be
a convective star, and suggest that it could therefore be the true
origin of the X-ray emission from this object. In the cases of HK~Ori
and V380~Ori, we present new measurements of the relative positions of
already-known companions, indicating orbital motion.  For HK~Ori,
photometric measurements of the brightness of the individual
components in four bands allowed us to decompose the system spectral
energy distribution (SED) into the two separate component SEDs.  The
primary exhibits a strong infrared excess which suggests the presence
of circumstellar material, whereas the companion can be modelled as a
naked photosphere.  
The infrared excess of HK~Ori~A
was found to contribute around two thirds of the total emission from
this component, suggesting that accretion power contributes
significantly to the flux.
Submillimetre constraints
mean that the circumstellar disk cannot be particularly massive, whilst the
near-infrared data indicates a high accretion rate. Either the disk
lifetime is very short, or the disk must be seen in an outburst phase.
 }

   \maketitle
%

\section{Introduction}

Herbig Ae/Be stars (HAeBes) are young intermediate-mass
(2--8$M_{\odot}$) pre-main-sequence stars first defined by Herbig
(1960). According to his definition, HAeBes should have emission lines
in their spectra, an early spectral type and be associated with dark
cloud material as an indication of their youth.  Despite the
similarities between HAeBes and the lower-mass T Tauri stars,
questions such as the existence of circumstellar disks, the
multiplicity and clustering of the objects, the origin of X-ray
emission (Zinnecker \& Preibisch, 1995), and the outflow activity are
much less clarified in the case of HAeBes.

The frequency and properties of binary systems amongst all stars are
of great importance in understanding the star formation process, since
they can point to certain star formation scenarios and rule out or
strongly constrain others. The configuration of pre-main-sequence
multiples is of special importance because the binary properties of
the objects can be compared with their formation environment, for
example whether in a T association or massive star forming region.
There is also less scope for the system configuration to have been
altered by stellar interactions or continuing accretion. The angular
momentum of the binary can also be compared to that of circumstellar
or circumbinary disks which may still exist at this stage.  The
coplanarity of the system orbit, the individual disks of the
components and a possible circumbinary disk are all major constraints
on viable formation mechanisms.

\begin{table*}[ht]
\caption{Summary of the observations. For each observation, the epoch
and observing wavelength is shown, together with the recovered binary
separation (in both mas and AU for the assumed distance), position
angle (measured anticlockwise from north), and flux ratio (typical
uncertainty 10--20\%). Where photometric calibration was performed,
the individual magnitudes of the components are also given.}
\label{postable}
\begin{center}
\begin{tabular}{ccccccccc}
\hline \hline
\multicolumn{7}{l}{{\bf \lkha}} \\
\hline
Epoch   & $\lambda_c$ & $\Delta\lambda$ & $\rho$     & $\rho$ & PA         & $m_A$ & $m_B$ &  $F_s/F_p$ \\
        & [nm]        & [nm]            &   mas        &   AU     &            &                             \\
1996.75 & 2191        & 411             &  66.3 $\pm$4 & 39.8     & 242 $\pm$2 &                            &       &     0.25         \\
1997.79 & 2165        & 328             &  66.6 $\pm$3 & 40.0     & 242 $\pm$2 &                            &       &     0.20             \\
1998.45 & 2110        & 192             &  65.8 $\pm$3 & 39.5     & 245 $\pm$2 &                            &       &     0.20          \\
1999.74 & 1648        & 317             &  63.9 $\pm$2 & 38.3     & 246 $\pm$1 &                            &       &     0.17        \\
2001.83 & 2115        & 214             &  58.4 $\pm$3 & 35.0     & 250 $\pm$1 &                            &       &     0.16           \\
2002.73 & 1648        & 317             &  50.4 $\pm$2 & 30.2     & 255 $\pm$2 &                            &       &     0.16          \\
2003.78 & 2115        & 214             &  48.6 $\pm$3 & 29.2     & 253 $\pm$2 &                            &       &     0.27           \\
\hline
\multicolumn{7}{l}{{\bf Elias~1}} \\
\hline
1996.75 & 2191        & 411                &  50.5 $\pm$4 & 7.07     & 54 $\pm$3 &                          &       &     0.69     \\ 
2003.76 & 2115        & 214                &  60.4 $\pm$1 & 8.46     & 59 $\pm$1 &                          &       &     1.00        \\
\hline
\multicolumn{7}{l}{{\bf HK~Ori}} \\
\hline
1995.18 &  550        &  60             &  334.2 $\pm$11.5  & 153.72        & 44.1 $\pm$1.5 & 12.13$\pm$0.2 & 13.01$\pm$0.2 & 0.45  \\  
1995.18 &  656        &  60             &  333.4 $\pm$11.5  & 153.35        & 44.6 $\pm$1.5 & 11.62$\pm$0.2 & 12.53$\pm$0.2 & 0.43   \\
1997.80 &  1613       &  304            &  339.2 $\pm$6.2   & 156.03        & 43.8 $\pm$0.5 & 8.55$\pm$0.14 & 9.76$\pm$0.16 & 0.33    \\
1997.80 &  2165       &  328            &  343.1 $\pm$6.8   & 157.83        & 44.2 $\pm$0.5 & 7.42$\pm$0.16 & 9.65$\pm$0.17 & 0.13   \\
2003.78 &  2115       &  214            &  347.7 $\pm$2.5   & 159.95        & 41.8 $\pm$0.7 &               &               & 0.17    \\
\hline
\multicolumn{7}{l}{{\bf V380~Ori}} \\
\hline
2003.78 &  2115       &  214            &  122.7 $\pm$2.5   & 56.44         & 224.0 $\pm$2.0 &              &               & 0.27   \\ 
\hline
\end{tabular}
\end{center}
\end{table*}

Numerous studies using a variety of techniques have established that
the majority of low-mass young stars are members of binary systems
(e.g. Ghez et al. 1993, Leinert et al. 1993, Simon et al. 1995).  The
T~Tauri population in the nearby Taurus-Auriga T-association displays
a factor of 2 higher binary fraction than that seen amongst
main-sequence stars.  At the high end of the stellar mass range,
bispectrum speckle interferometry of the massive stars in the Orion
nebula cluster revealed companions for 7 out of 13 target objects,
which after correction for undetected systems was estimated to
represent a binary frequency close to 100\%, with an average of $>1.5$
companions per primary (Preibisch et al. 1999). This is about three
times the multiplicity of low-mass stars, and may suggest an
alternative formation scenario for the high mass objects. Studies
aimed specifically at intermediate mass stars have been rare. Li et
al. (1994) carried out an infrared imaging study of 16 sources and
reported companions for nine of them, although of these many were
widely separated and are probably not bound.  Leinert et al. (1997,
hereafter LRH97) studied a larger sample of 31 stars using a speckle
interferometry technique in the near-infrared, and found that 31\% of
the systems were multiple with separations between 50 and
1300~AU. This was estimated to represent an excess binarity among
HAeBe stars compared to main-sequence G dwarfs of a factor of
two. Pirzkal et al. (1997) used a speckle shift and add technique to
search for companions around 39 HAeBe stars. They detected 9 multiple
systems, and estimated on statistical grounds that this implied a true
binary frequency amongst HAeBes of at least 85\%.  A survey for
spectroscopic binaries amongst HAeBe stars by Corporon \& Lagrange
(1999) found a binary frequency of 17\%. For short periods ($<$100
days) the binary frequency was found to be 10\%, compared to the
frequency amongst WTTS spectroscopic binaries of $ 11 \pm 4$\%
(Mathieu 1992) or the frequency amongst main sequence solar-mass stars
of $ 7 \pm 2$\% (Duquennoy \& Mayor 1991).

In this paper, we present bispectrum speckle interferometry at
multiple epochs of four HAeBe systems.  For two of these, \lkha\ and
Elias~1, our diffraction-limited images show unambiguously for the
first time that they have close binary companions.  For the known
binary HK~Ori, we separate the SEDs of the two companions and by
considering the different colours show that the brighter object
possesses a circumstellar disk which the secondary lacks.  For
V380~Ori, we are able to show relative motion compared to the position
measured by LRH97.

\section{Observations and data reduction}
\label{observations}

\begin{figure*}[!ht]
   \psfig{figure=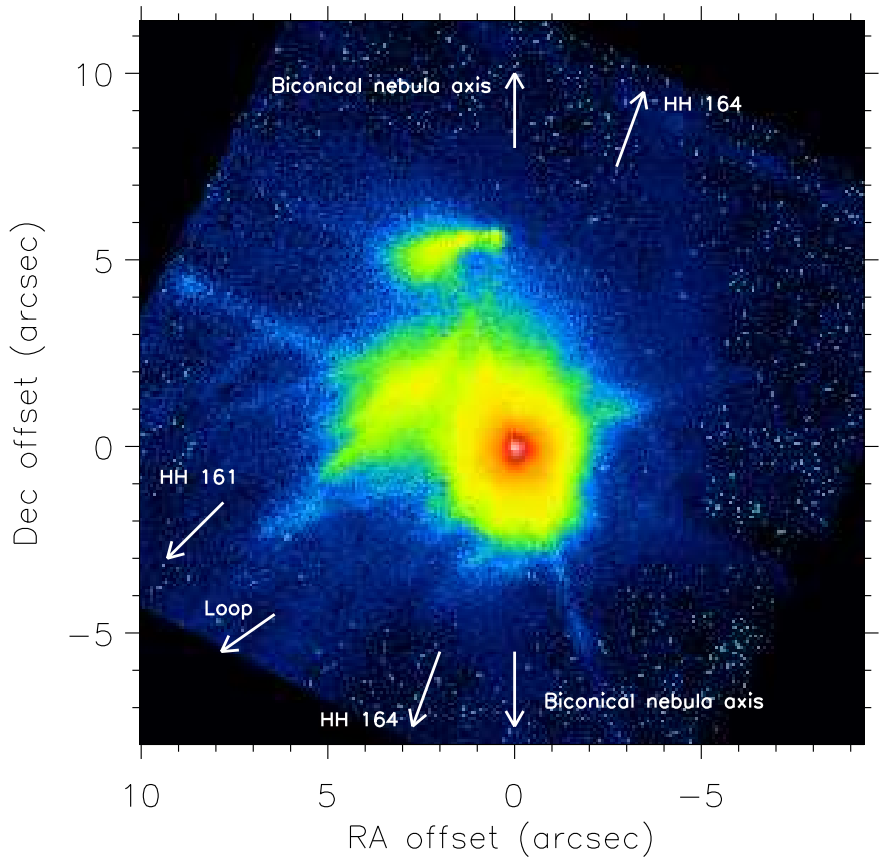,width=12.0truecm} 
{\caption{HST image
  of \lkha, taken with the NICMOS camera at 1.1$\mu$m. North is up and
  East to the left. The source approximately 6$''$ due north of \lkha\
  is the already known companion \lkha~B. A fan-shaped trail of emission
  is visible to the east of this companion.  The new speckle 
companion, \lkha~A2, lies within
  60~mas of the primary and is not discernable in this image (see
  Fig.~\ref{lkhaclose}).  The directions to various associated HH
  objects and other structures are indicated. The various structures
  comprising HH~164 were identified by Corcoran et al. (1995) as the
  jet of \lkha. The HH object HH~161 is believed to be powered by the
  companion \lkha~B, and the arrow pointing to this object is shown
  radiating from the position of this object. Interesting streamer-like
structure can be seen to the east of \lkha\ with the same 
position angle as HH~161, perhaps indicating that the 
HH~161 flow interacts with the material around \lkha. The axis of the
  biconical nebula mapped by Perrin et al. (2004) is indicated,
  running nearly exactly North-South. The opening angle of this structure is
  approximately 30$^{\circ}$, and so the outflow HH~164 emerges from
  within the supposed bipolar cavity, even though its direction is not
  exactly aligned with the axis.  }
           \label{lkha198pic}}
\end{figure*}

The optical observations of HK~Ori were made with the 2.2m ESO/MPG
telescope on March 8, 1995.  The speckle interferograms were recorded
through interference filters with central wavelength/bandwidth of
550\,nm/60\,nm and 656\,nm/60\,nm.  The detector used for the visible
observations was an image intensifier (gain 500\,000) coupled
optically to a fast CCD camera (512$^2$ pixels/frame, frame rate 4
frames/s).  All the other speckle
interferograms were recorded with the SAO 6\,m telescope in Russia
between 1995 and 2003. The detector of our speckle camera at SAO was
either a Rockwell HAWAII array (1998-2003; only one 512x512 Quadrant
was used) or a 256x256 PICNIC array detector (1996 and 1997). The
visible data sets consist of about 700 to 1400 speckle interferograms
each with an exposure time of 50~msec.  The infrared sets consist of
between 400 and 3200 speckle interferograms each with an exposure
time of between 150 and 250~msec.

The object power spectra were determined with the speckle
interferometry method (Labeyrie 1970). Speckle interferograms of
unresolved single stars were recorded just before and after the object
and served as references to determine the speckle transfer
function.  Diffraction-limited images were reconstructed using the
bispectrum speckle interferometry method (Weigelt 1977; Weigelt \&
Wirnitzer 1983; Lohmann et al. 1983; Hofmann \& Weigelt 1986).

For HK~Ori, photometric calibration in the K$'$-band was carried out by
observing the photometric standard stars HD~18881 and Gl~105.5, chosen
from Elias et al. (1982), on October 19, 1997. 
For the photometric calibration in the H-band the
photometric standard star HD~40335 and the star HD~31648 were observed
on October 22 and 19, 1997. The
photometric calibration in the visible is based on our speckle
observations of NX~Pup from March 10, 1995 (Sch\"oller et
al. 1996).  Simultaneously to our observations of NX~Pup, CCD
photometry of the unresolved pair NX~Pup\,A/B was carried out at the
Danish 1.5m telescope at La Silla.  Observations of standard stars
taken from the list by Landolt (1992) allowed for the absolute
photometric calibration.

\section{Results}
\label{results}

Each of the observed visibility functions clearly showed a pattern
suggestive of a binary system. The separation and position angle of
the system were measured directly from the two dimensional visibility
function. Maps of the objects were made and the flux ratios were also
measured. The recovered parameters for the various systems are listed
in Table~\ref{postable}. Below, we discuss each object individually.

\subsection{\lkha}

LkH$\alpha$~198 (SIMBAD coordinates: $\alpha$= 00$^h$~11$^m$~25.97$^s$,
$\delta$=+58$^{\circ}$~49$'$~29.1$''$, J2000) is a well-studied HAeBe
located in the dark cloud Lynds~1265 at a distance of about 600~pc.
It is associated with another well-known HAeBe star, V376~Cas, which
lies approximately 35$''$ to the north. A schematic picture of the
complex environment around these two sources is given in the paper by
Koresko et al. (1997). We reproduce a previously unpublished HST image from the archives in
Fig.~\ref{lkha198pic}, which shows the inner $15''\times15''$ of the system.
Other images of the inner region of the system are shown in 
Koresko et al. (1997), Fukagawa et al. (2002) and Perrin et al. (2004).

Lagage et al. (1993) detected an embedded source (\lkha~B) 6$''$ north
of \lkha in 10~\mic\ images which is also visible at 2.2~\mic\ (Li et
al. 1994) and in the 1.1~\mic\ HST image shown here.  A deeply
embedded protostar (\object{\lkha~MM}) was found about 19$''$ to the
northwest by Sandell \& Weintraub (1994) using the JCMT at 800~\mic.
This object dominates in the millimetre and submillimetre but is not
visible at all at shorter wavelengths. The most striking extended
feature in the optical is a large bubble-like elliptical loop
extending some 40$''$ to the southeast at a position angle of
approximately 135$^{\circ}$ (Corcoran et al. 1995).  Coronagraphic
imaging in the near-infrared by Fukagawa~et~al.  (2002) showed in more
detail the close environment of \lkha. Their image showed the primary
extended along a roughly north-south axis (PA=15$^{\circ}$).  This
elongation was also reported by Koresko~et~al. (1997), and is visible
in the HST image in Fig.~\ref{lkha198pic}, but here the issue is
complicated by the presence of a diffraction spike along the same
axis.

Numerous HH objects are found around \lkha\ (see e.g. Corcoran et
al. 1995, Molinari et al. 1993, Goodrich 1993). Corcoran et al. (1995)
found knots of SII emission tracing both sides of a jet from \lkha\ at
a position angle of around 160$^{\circ}$. a further knot of emission
was seen along the same axis on the opposite side of \lkha\ at
position angle 340$^{\circ}$.  These objects have been designated
\object{HH~164}.  Spectra of these emission knots revealed that they
have a low radial velocity, and therefore suggest that the jet lies
close to the plane of the sky. Another knot seems to trace the
southeast part of a jet apparently originating from the companion
\lkha~B along the axis of the loop. This knot, together with [SII]
emission at the loop apex, has been designated \object{HH~161}. The
trail of nebulosity that can be seen extending eastwards from the
previously-known companion \lkha~B may be the northern edge of an
outflow cavity caused by the flow from this source driving
HH~161. Nebulosity is also seen extending east from the main object.
This region contains ``streamer''-like structure with the same
position angle as HH~161 and this might indicate interaction of the
outflow from \lkha~B with material surrounding the primary. Aspin \&
Reipurth (2000) discovered a further HH object, \object{HH~461}, with
a faint bowshock morphology, along the axis of the \lkha\ jet some
80$''$ to the southeast. McGroarty \& Ray (2003) identified two
further bowshock-shaped structures, designated \object{HH~801} and
\object{HH~802}, which lie along the line of HH~164 with a total
separation of 2.3~pc.
\begin{figure}[!h]
\vbox{
\begin{center}
     \psfig{figure=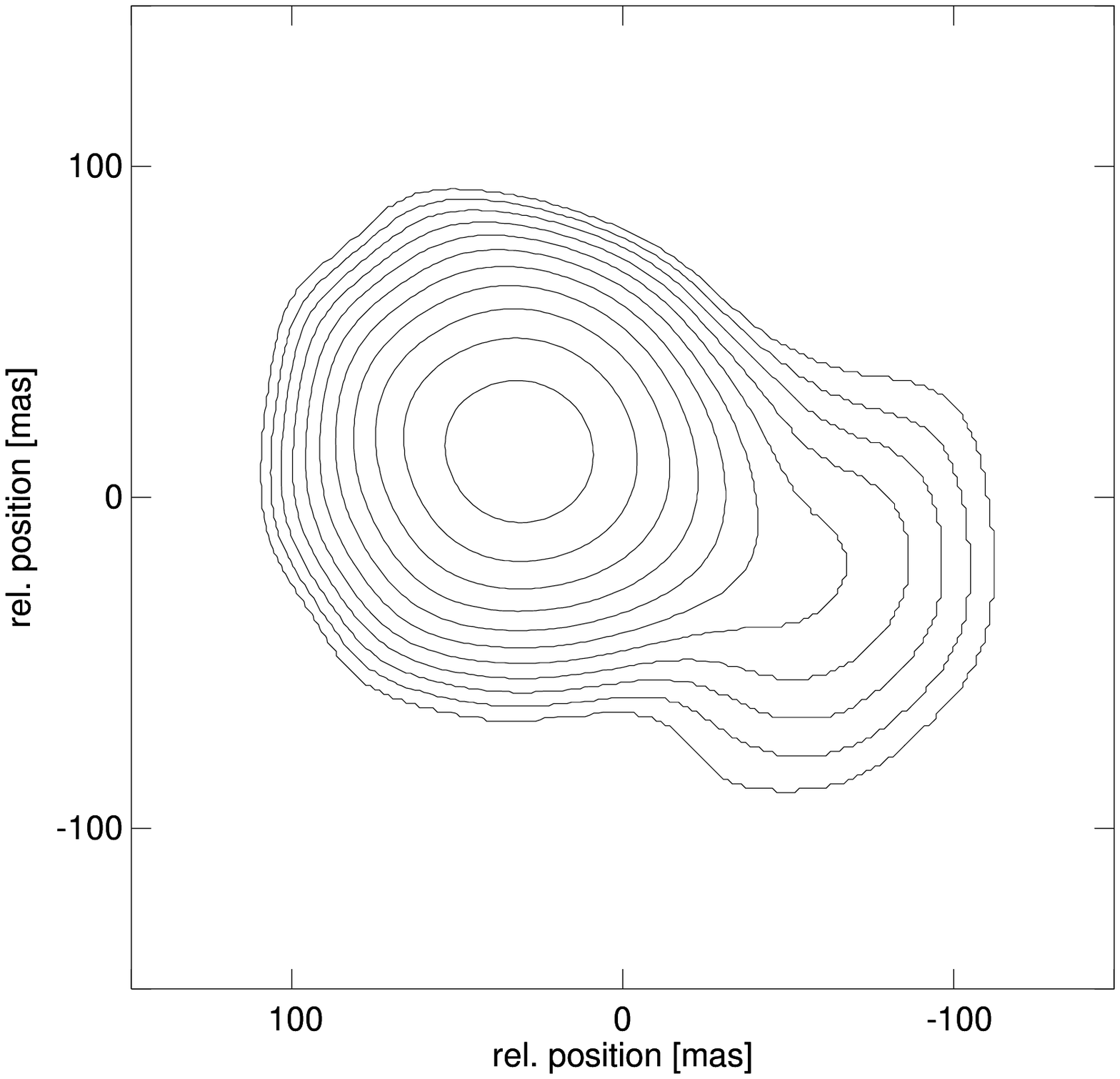,width=6.0truecm} 
   \psfig{figure=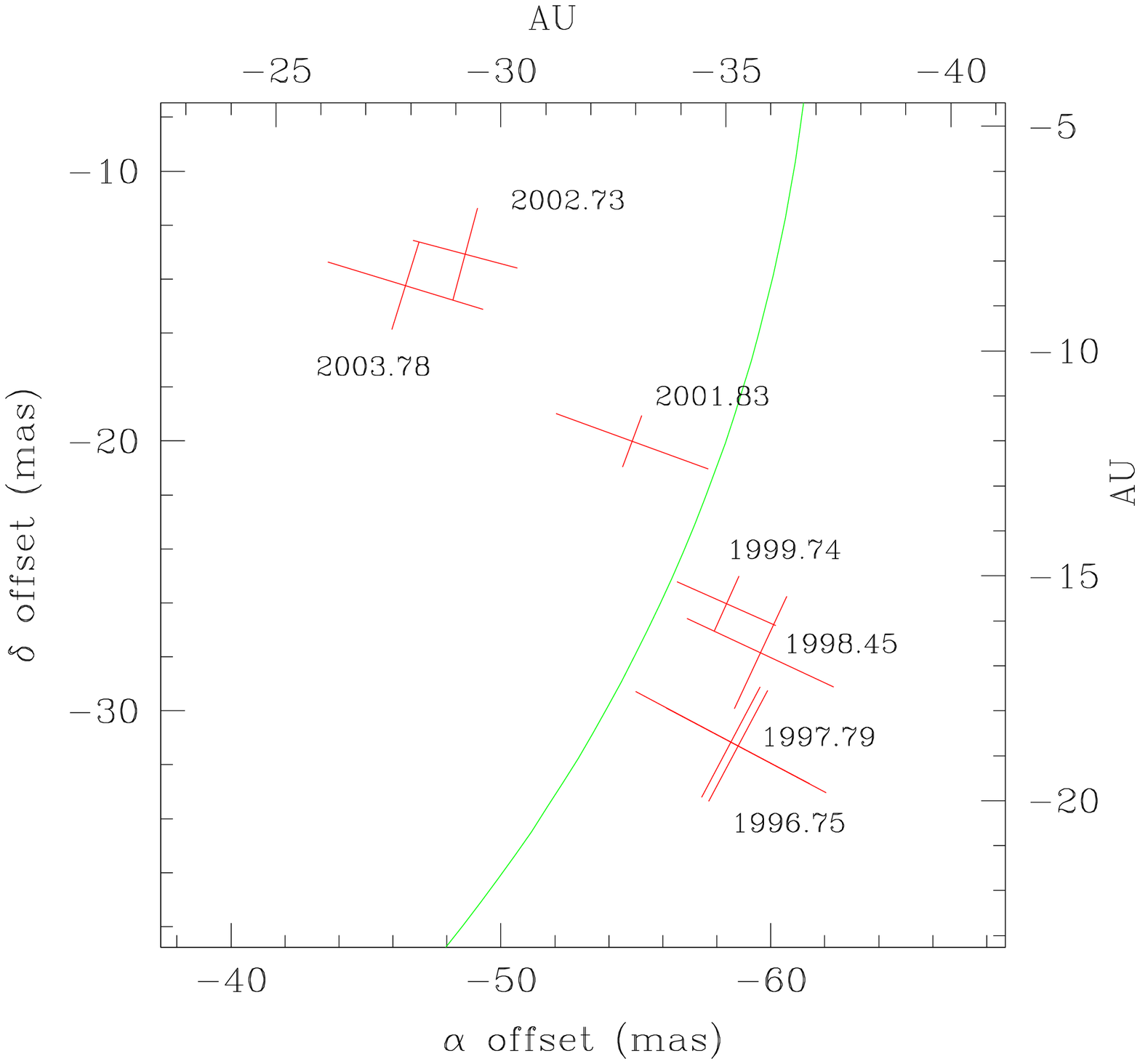,width=7.0truecm}
\end{center}
} {\caption{Top: Contour map of the \lkha\ system constructed from
  speckle observations on the 26th October 1999.  Bottom: The position
  of the \lkha\ companion relative to the primary. The scale is given
  in mas and AU for an assumed distance of 600~pc. The solid line is
  a circle of radius 37~AU, for an assumed distance of 600~pc, centred
  on the primary}
           \label{lkhaclose}}
\end{figure}

Polarization maps in the {\em I} band by Leinert et al. (1991) and Piirola
et al. (1992), and at {\em V} and {\em I} by Asselin et al. (1996) all
show a centrosymmetric polarization pattern extending in an arc to the
southeast.  The map of Piirola et al. (1992), which was taken under
conditions of extremely good seeing, also shows a similar 
centrosymmetric pattern extending to the northwest. A region of
polarization perpendicular to the centrosymmetric pattern was seen
within the central 0\farcs5.  This was taken to be the signature of a
central, unresolved disk. 
Perrin et al (2004) used laser guide star
adaptive optics to obtain very high quality near-infrared $J$, $H$ and $K_s$ band
polarimetry of the central part of the \lkha\ system.  They saw a
polarized biconical nebula aligned almost exactly north-south, with a
dark unpolarized central region. This biconical nebula, which can be
interpreted as scattering from an evacuated bipolar cavity caused by
an outflow, is consistent with \lkha\ being the source of the outflow
HH164, but the axis is apparently misaligned by approximately
20$^{\circ}$.

Our speckle observations of \lkha\ span seven years, from 1996.75 to
2003.78. The two dimensional visibilities show clear signatures of
binarity, and the position of the companion, which we designate
\lkha~A2, relative to the primary (\lkha~A1) changes significantly
over the period of the observations.  The track of the secondary
relative to the primary is shown in Fig.~\ref{lkhaclose}. The
trajectory shows signs of curvature to the east, indicating that we
are seeing a segment of an orbit, although it appears a straight line
trajectory cannot be excluded. A map of the system reconstructed from the 
1999 observations, is also shown in Fig.~\ref{lkhaclose}.

\begin{figure*}[!ht]
\vbox{ \hbox{ \psfig{figure=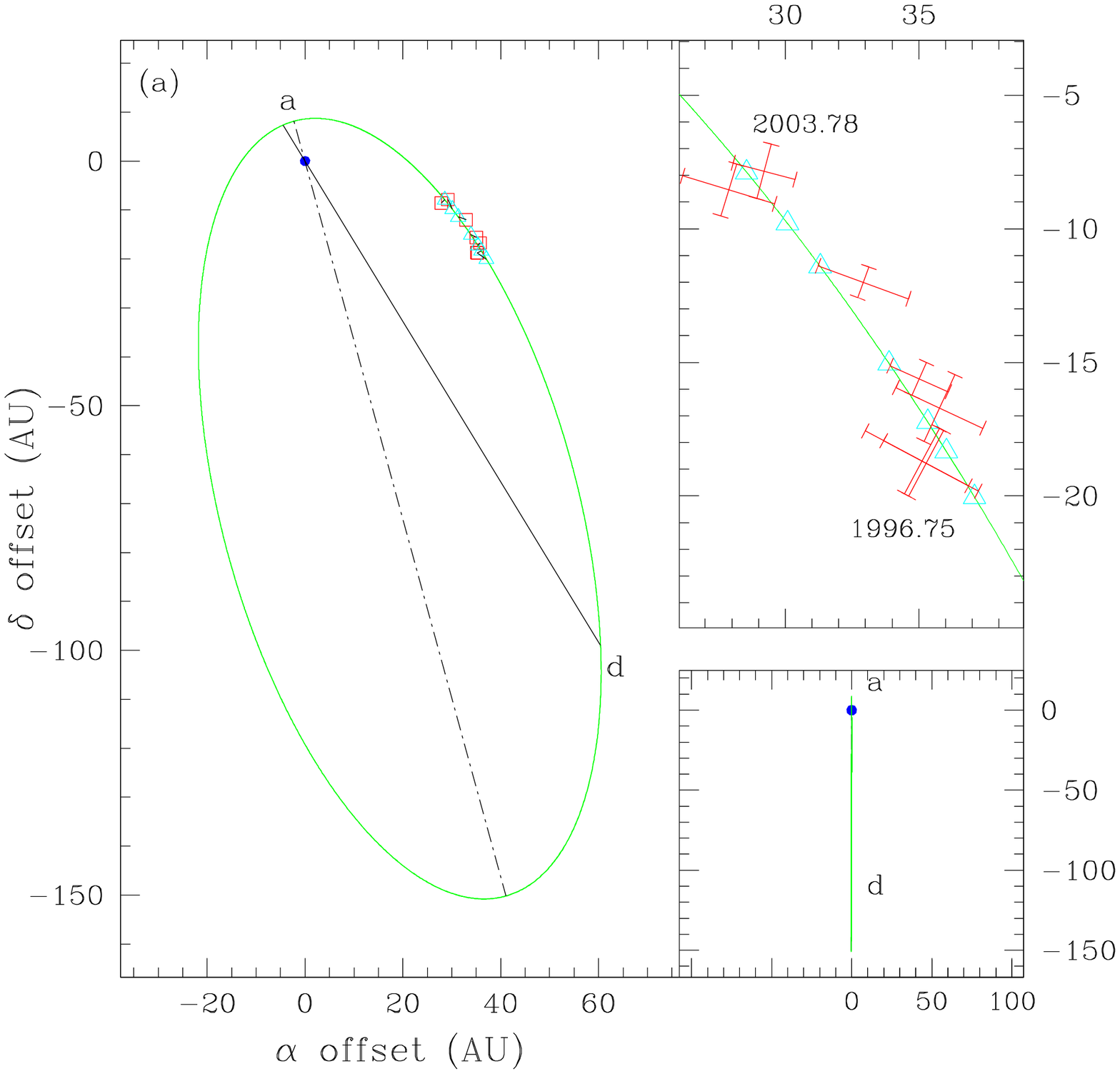,width=8.0truecm}
\hfill \psfig{figure=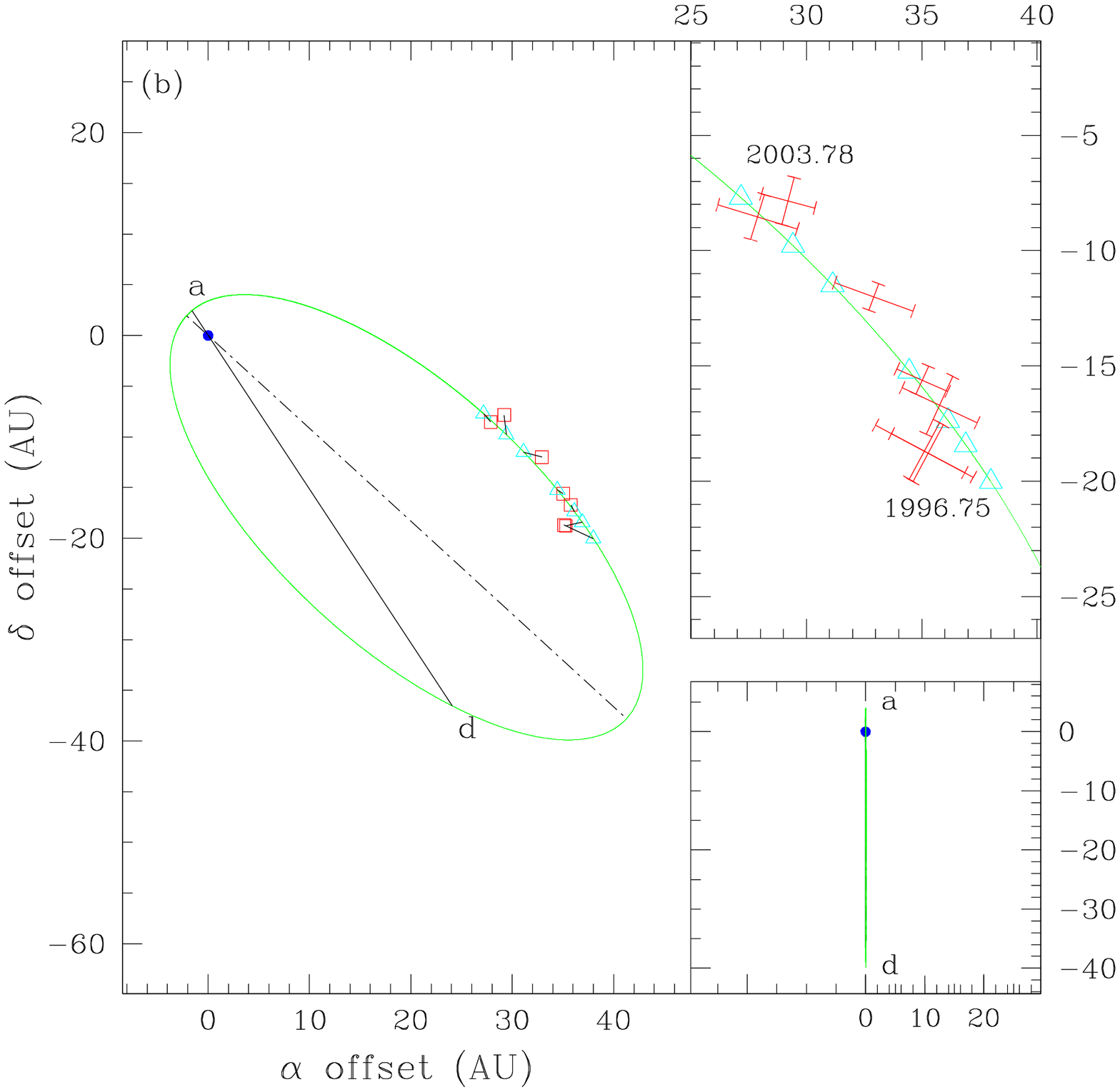,width=8.0truecm} \hfill }
\hbox{ \psfig{figure=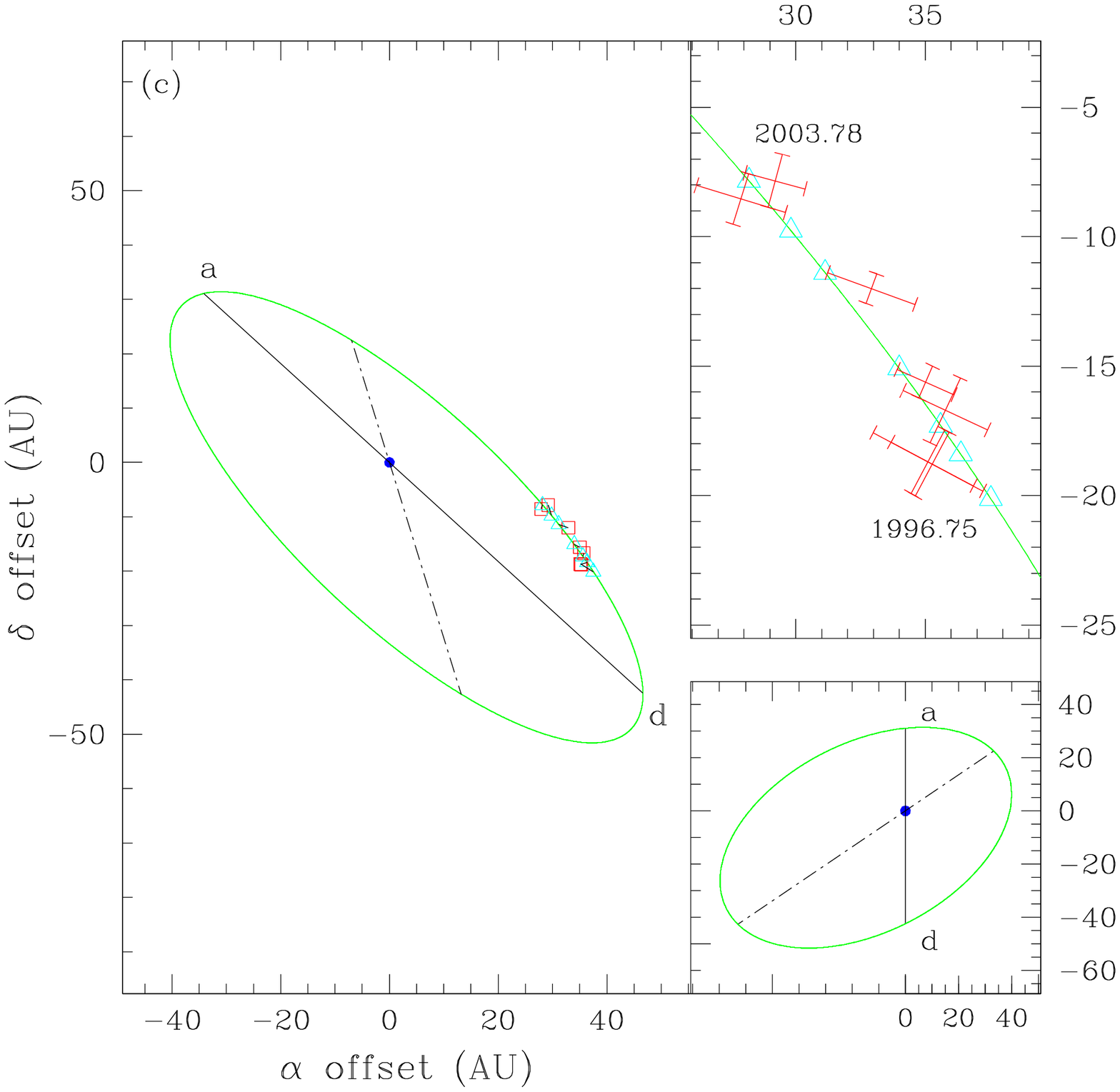,width=8.0truecm} \hfill
\psfig{figure=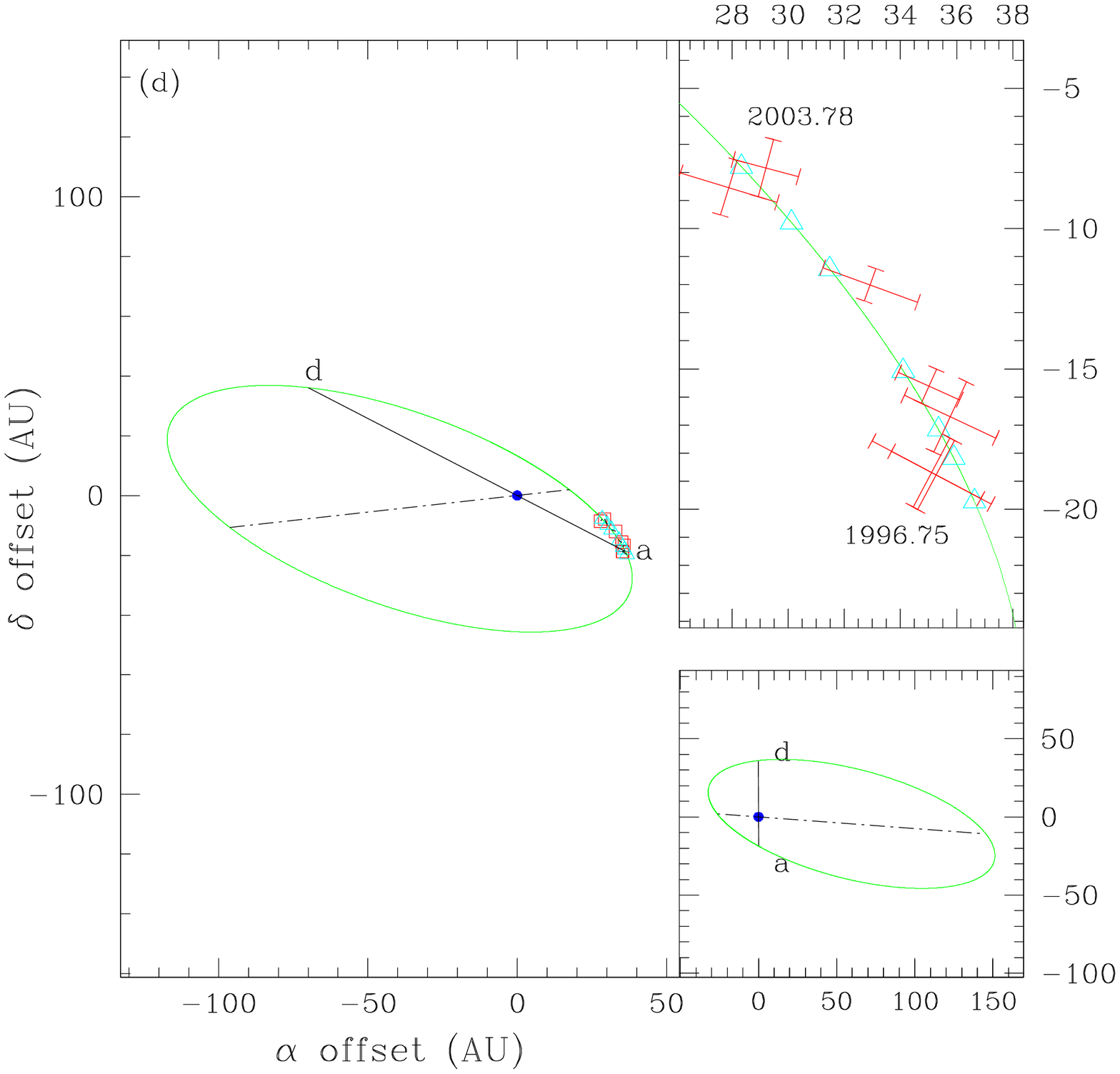,width=8.0truecm} \hfill }}
{\caption{\label{orbits} Orbit fits to the motion of the
LkH$\alpha$~198 companion for various choices of initial
parameters. The main panel of each plot shows the fitted orbit
projected on the plane of the sky. North is up and east to the
left. The measured data points are plotted as open squares and the
corresponding points in the fit are plotted with open triangles.  A
dash-dotted line marks the major axis, and a solid line marks the line
of nodes (i.e. the line of intersection between the orbital plane and
the plane of the sky). The ascending and descending nodes are labelled
'a' and 'd'. In the inset at upper right in each figure, we show the
segment of the orbit with the data points on a larger scale.  The data
points are shown with errorbars, and the fitted points are again
plotted as open triangles. The first and last
points are labelled with the observing epoch.  In the inset at lower
right, we show the projection in a plane perpendicular to the plane of
the sky. North is up and the radial direction (pointing away from the
observer) is to the left. The coordinate systems are indicated in each
panel of the top left-hand plot. {\bf (a)}~Fit for an assumed primary
mass of 1.6$M_{\odot}$ and secondary mass 1.0$M_{\odot}$. The initial
guess for the fit involves a low velocity in the $R$ direction and a
small displacement from the plane of the sky. As a result, the fitted
orbit lies close to the plane of the sky.  {\bf (b)}~The same low
starting values for $R$ position and velocity, but for an assumed
primary mass of 4.0$M_{\odot}$ and secondary mass 2.5$M_{\odot}$.
{\bf (c)}~A fit with an assumed primary mass of 4.0$M_{\odot}$ and
secondary mass 2.5$M_{\odot}$, with starting conditions chosen to
produce a near-circular orbit. {\bf (d)} This fit has masses of 4 and
2.5$M_{\odot}$ and a high initial radial velocity, which is intended
to produce an orbital plane as highly inclined as possible to the
plane of the sky. }}
     \end{figure*}

\subsubsection{Orbital motion}

With only seven points,
spanning a short segment of the apparent orbit, it is not possible to
determine the true orbit with any certainty. Nevertheless, we make
some preliminary orbital fits and explore the constraints which can be
imposed on the motion.

The exact configuration of the orbit is particularly interesting in
the case of \lkha\ because of the complex environment of the
system. The outflow has a low radial velocity, suggesting that it lies
in the plane of the sky and that the inner disk should be seen nearly
edge-on. This is also the impression given by the polarization maps of
Perrin et al. (2004).  Long-term photometric monitoring by Mel'nikov
(1997) shows that \lkha\ lacks the strong photometric variations which
define UXOR sources, despite the supposed edge-on viewing angle of any
disk.  According to the prevailing theories of UXOR variability, in
which the deep photometric minima are caused by obscuration by clumps
in a disk, this makes the object something of an anomaly. The presence
of the companion and the details of the orbit are interesting in the
context of the complex environment, and we therefore make a more detailed discussion of the orbit
than would be justified for an isolated system with such a short known
orbital segment.
\begin{table*}[!ht]
\caption{Parameters of the various fitted orbits shown in
Fig.~\ref{orbits}.  The parameters shown are the mass of the primary,
mass of the secondary, major axis length, eccentricity, period,
position angle of the ascending node (measured anticlockwise from
north in the plane of the sky), inclination between the orbital plane
and the plane of the sky, and the longitude of periastron, defined as
the angle between the ascending node and the periastron measured in
the plane of the true orbit and taken in the direction of the
secondary's motion. All the orbits appear prograde on the sky. The
last line, labelled ``Straight Line'', shows the $\chi^{2}$ for the
best straight-line fit.}
\begin{center}
\begin{tabular}{rrrrrrrrrr}
\hline 
\hline 
Fit & $M_{1}$($M_{\odot}$)& $M_{2}$($M_{\odot}$)  &Major axis (AU) & $e$ & $P$(yr)    &  $\Omega$ & $i$ & $\omega$ & $\chi^{2}$ \\
\hline
\hline
(a)          & 1.6 & 1.0 & 164.2 & 0.90 & 462.7 & 31  & 0.3  & 344 &  1.32 \\
(b)          & 4.0 & 2.5 & 58.8  & 0.90 & 62.7  & 33  & 0.4  & 14  &  1.37  \\
(c)          & 4.0 & 2.5 & 117.9 & 0.31 & 178.0 & 48  & 70   & 300 &  1.34 \\ 
(d)          & 4.0 & 2.5 & 206.3 & 0.69 & 412.1 & 243 & 70   & 63  &  1.15 \\
Straight Line&     &     &       &      &       &     &      &     &  1.56 \\
\hline
\hline
\label{orbitstab}
\end{tabular}
\end{center}
\end{table*}

We made a number of fits to the observed relative motion of the \lkha\
companion. The fits were made numerically by running a 2-body code
multiple times and minimizing the resulting deviations from the
points.  Conventional orbital parameters for the best fit, such as the
major axis, eccentricity and so on, can be recovered from a full
simulation of the best-fit orbit.

The mass of the primary was estimated by Hillenbrand et al. (1992) to
be 1.6~M$_{\odot}$ based on the luminosity estimated from the
dereddened $V$ magnitude and a spectral type of A5. However, Natta et
al.  (1992) estimated the luminosity to be 250~$L_{\odot}$ from far
infrared measurements, which by comparison with pre-main-sequence tracks 
from Palla \& Stahler (1993), suggests a mass of 3.5--4$M_{\odot}$, if
the bulk of this luminosity can be assumed to originate from the
primary. If half this luminosity arises from accretion power, the
estimated primary mass would fall to 3.0-3.5$M_{\odot}$.  For an
orbit calculation, to the mass of the primary must be added the mass
of any compact circumstellar structure such as a disk.  Leinert et
al.'s (1991) estimate of the mass of the immediate circumstellar
material was only 0.01~$M_{\odot}$, and so we neglect the possible
mass of circumprimary and circumsecondary material.  The flux ratio of
the secondary to the primary is measured to be 0.20 at $K'$ and 0.16
at $H$ (see Table~\ref{postable}). If this is taken as a rough
estimate of the luminosity ratio, the mass of the secondary is
estimated {\em from evolutionary tracks} 
to be either approximately 1.0~$M_{\odot}$ if the primary is
1.6~$M_{\odot}$ or 2-2.5~$M_{\odot}$ if the primary is around
3.5-4~$M_{\odot}$.  For the sake of fitting the orbit, we have adopted
two different masses for the system, one of 1.6 and 1.0 $M_{\odot}$,
following Hillenbrand et al. (1992), and one of 4.0 and 2.5
$M_{\odot}$, following Natta et al. (1992).

The motion on the sky leads to tight constraints on the motion in the
plane of the sky.  Because such a short segment of the orbit is
observed, however, there is little constraint on the possible initial
position and velocity in the radial direction. On the sky, the
secondary covers an apparent distance of almost 12 AU in a time of six
years. This implies a velocity in the plane of the sky of 9.5~km/s.
For a system comprising a 1.6$M_{\odot}$ and 1.0$M_{\odot}$
components, the escape velocity would be approximately 11~km/s.  This
implies that, if the system has such a low mass, the orbit must lie
close to the plane of the sky (assuming it is in fact bound). On the
other hand, for a system with a total mass of 6.5$M_{\odot}$, the
escape velocity is approximately 17~km/s, so that the true velocity
can be substantially larger than the apparent velocity on the plane of
the sky, and the orbit can be significantly inclined to the line of
sight.

In Fig.~\ref{orbits}, we show four fitted orbits for various assumed
system parameters. The orbital parameters and fit $\chi^{2}$ values
for these are listed in Table~\ref{orbitstab}.  The top left panel
shows the fit for a system with masses 1.6/1.0~$M_{\odot}$, and low
values for the initial value (i.e. before the fitting procedure) of
the initial radial position and velocity. This leads to a quite
eccentric orbit almost in the plane of the sky.  In the top right
panel, we show a fitted orbit with the same low values for the initial
radial position and velocity, for a system with masses
4/2.5~$M_{\odot}$. The lower left panel shows a fit for system with
4/2.5~$M_{\odot}$, for which the initial radial offset and velocity
were chosen to attempt to produce a near-circular orbit. In fact, the
eventual fit has an eccentricity of about 0.3.  Finally, a fit was
made with a very large initial radial velocity.  The intention of this
was to incline the fit as highly as possible to the plane of the sky,
so that it is as close as possible to coplanar with the supposed
edge-on disk at the heart of the system. The inclination of this fit
to the plane of the sky was approximately 70$^{\circ}$. This fit is
shown in the lower right panel of Fig.~\ref{orbits}.  Finally, a fit
was made with the secondary set a long way back from the plane of the
sky, and only the north-south and east-west motion varied. This
produces the best-fit straight-line motion, the $\chi^{2}$ of which is
listed in Table~\ref{orbitstab} for comparison with the bound fits.

\subsubsection{Results of orbit fitting}

Although the straight line has the highest $\chi^2$ (because it can't
account for the apparent curvature), there is no significant
difference between the quality of the various fits.  {\bf The most
edge-on orbit fits which could be made had an inclination of
approximately 20$^{\circ}$ to the plane of the sky, and the long axis
of the projected orbit was tilted by at least 20$^{\circ}$ to the
east-west direction.  This contrasts with the Perrin et al. (2004)
image which shows a disk-like structure aligned almost exactly
north-south, and the jet velocity measurements of Corcoran et
al. (1995) which show low velocities for the jet material and
therefore suggest that the jet lies very close to the plane of the
sky. Unfortunately, without tangential velocity measurements, it is
not possible to put firm limits on the inclination of the jet for
comparison with the model orbits. Interestingly, the orientation of
the axis of the projected orbit {\em (d)} is approximately aligned with the
direction of the HH~164 jet.  }

\subsection{Elias~1}
\begin{figure}[!h]
\vbox{
\begin{center}
  \psfig{figure=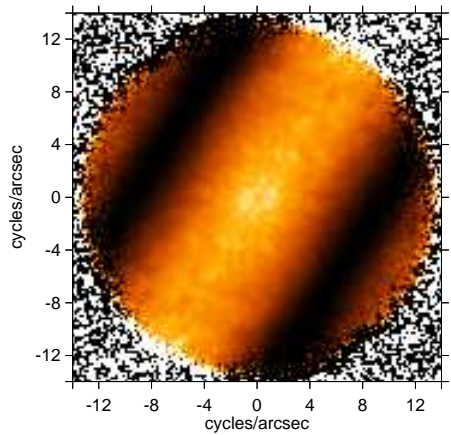,width=6.0truecm} 
  \psfig{figure=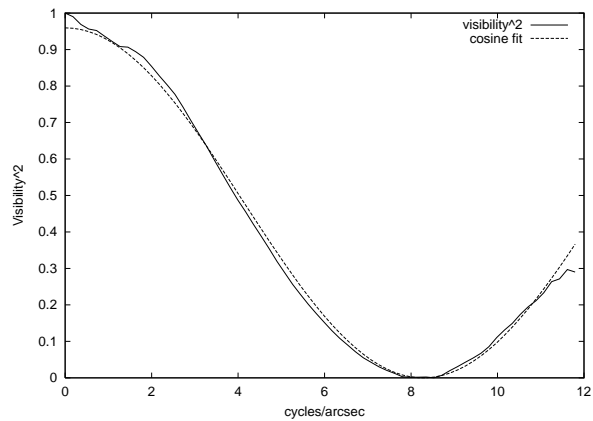,angle=270.,width=8.0truecm} 
\end{center}
}
  {\caption{Top: The 2d power spectrum of Elias~1 from 2003.76. 
Bottom: a cut across the fringes, with a cosine fit.}
           \label{elias1ps}}
\end{figure}

\begin{figure}[!b]
\vbox{
\begin{center}
  \psfig{figure=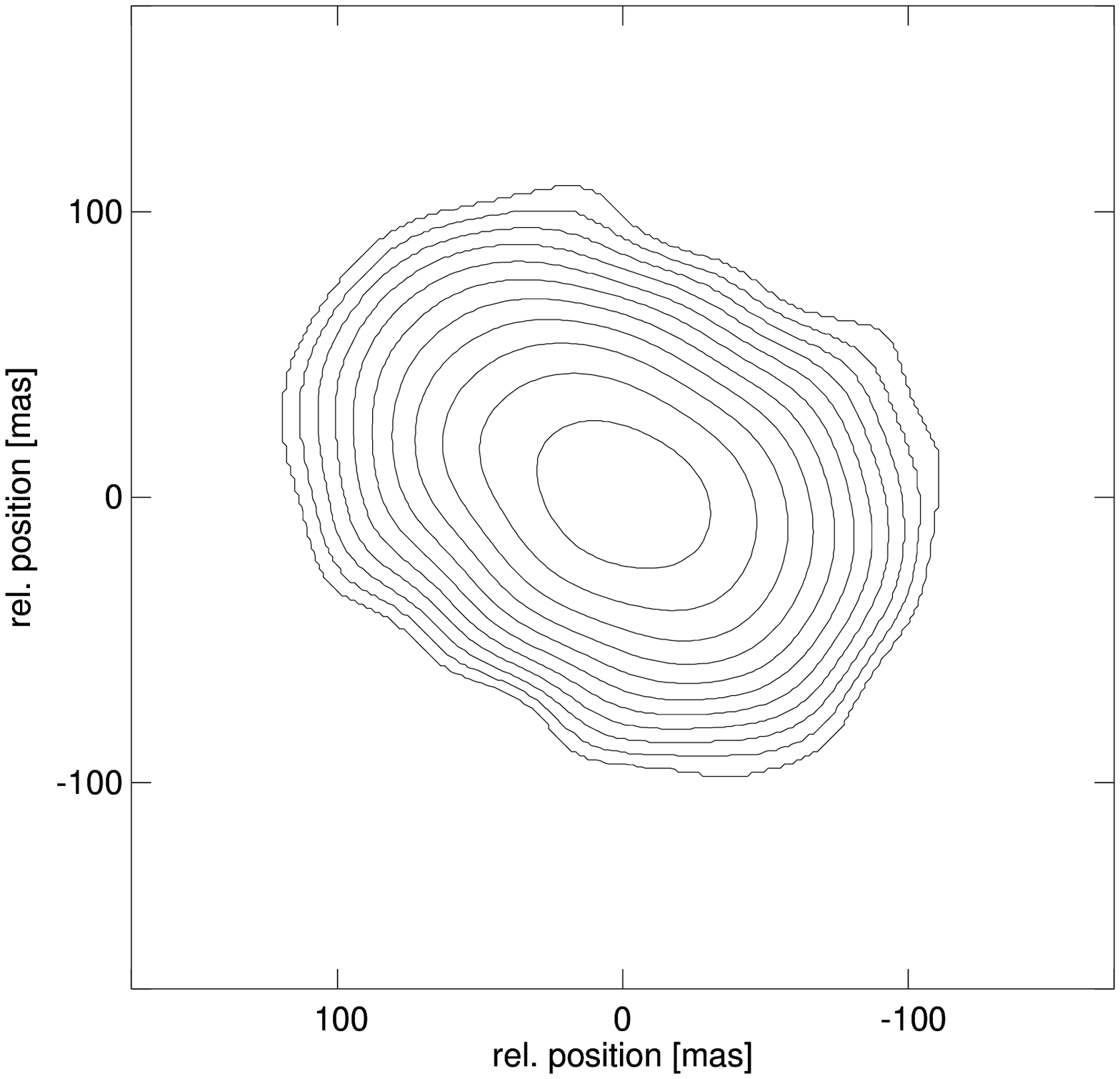,width=6.0truecm} 
   \psfig{figure=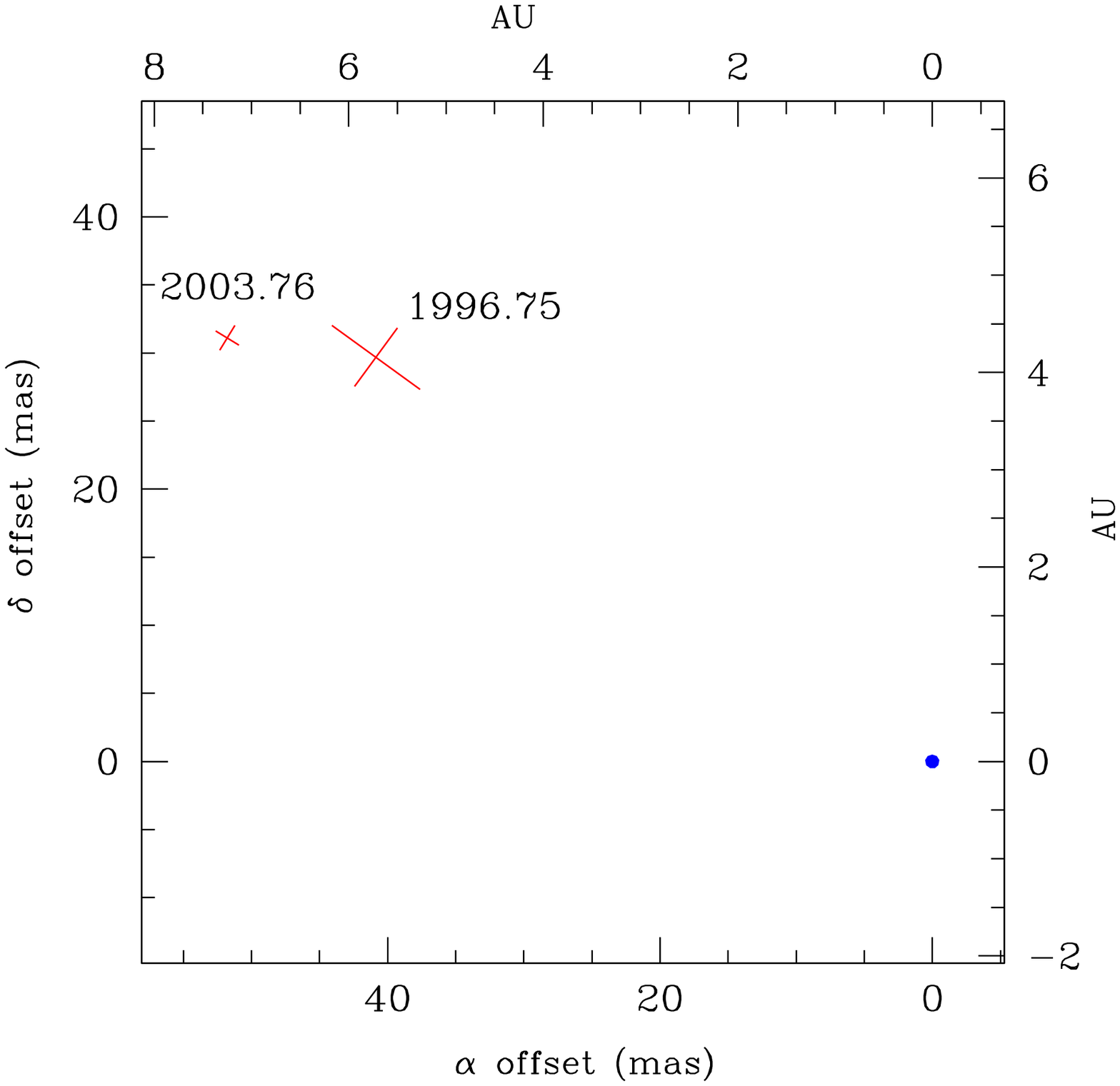,width=7.0truecm} 
\end{center}
}
  {\caption{Top: Contour map of Elias~1 made from the 2003 observations. 
Bottom: The position of the Elias~1 companion relative to 
the primary. The primary position is marked with a dot.}
           \label{elias1data}}
\end{figure}

Elias~1, also known as V892~Tau ($\alpha$= 04$^h$~18$^m$~40.60$^s$,
$\delta$=+28$^{\circ}$~19$'$~16.7$''$, J2000) is located in the
Taurus-Auriga complex at a distance of approximately 140~pc (Elias
1978).  Its spectral type was estimated as A6 with $A_V=3.9$~mag and
luminosity of around 38$L_{\odot}$ (Berrilli et al. 1992), but other
authors have assigned earlier spectral types of A0 (Elias, 1978) 
or B9 (Strom \& Strom 1994). Elias~1 was classed as a
type II HAeBe star by Hillenbrand et al. (1992), having a very flat or
even rising SED in the NIR to FIR. This suggests the presence of a
possible envelope as well as a circumstellar disk, or possibly a
flared disk. The polarization is 4.7\% (Yudin 2000).

A T Tauri type companion lies approximately 4$''$ to the northeast
(LRH97).  Kataza \& Maihara (1991) obtained one dimensional speckle
interferograms at {\em K} and {\em L}, with position angles of zero
and ninety degrees. They also obtained one-dimensional speckle
interferograms at {\em L}$'$ with PA=45$^{\circ}$, showing the object
was resolved in this direction. The object was found to be resolved in
the east-west direction, but unresolved in the north-south
direction. The preferred interpretation was a flattened halo
associated with a circumstellar accretion disk and extending some
40~AU in {\em K} or 100~AU in {\em L}. Haas et al. (1997) later
obtained 1-D specklegrams at {\em J},~{\em H},~{\em K} and {\em L}
with a larger 3.5m telescope. Their specklegrams were also oriented
north-south and east-west.  Mostly on the basis of the relative halo
brightness at {\em J} and {\em H}, they rejected the disk hypothesis
of Kataza \& Maihara (1991) and argued instead for a bipolar nebula
with east-west orientation.  They also considered a possible binary
interpretation, noting that a binary with position angle 45$^{\circ}$
or 135$^{\circ}$, separation $<0^{''}.1$ and brightness ratio 0.08 at
{\em K} or 0.2 at {\em L} was not ruled out by their data.

Elias~1 is unusual for HAeBe stars in being an X-ray source (Zinnecker
\& Preibisch 1994). Giardino et al. (2004) observed a strong X-ray
flare from the system, and also smaller X-ray variations which could
be identified as originating on Elias~1 and not the companion
Elias~1~NE.  Spectral fits with a two temperature model suggested a
hot component of around $kT \approx 2 - 3$~keV.  The time variability,
and the hot plasma temperature, point to magnetically confined plasma
in the neighbourhood of the star.  Elias~1 is also a radio source,
although its rising spectrum, though steep, is consistent with
thermal emission from a wind (Skinner et al. 1993).

Our data show that Elias~1 is a close binary with a separation of
approximately 50~mas and position angle approximately 50$^{\circ}$
(Fig.~\ref{elias1ps}).
Furthermore, the relative position of the secondary with respect to
the primary has changed in the seven years between the two observation
epochs. The positions and flux ratios are given in
Table~\ref{postable} and are shown in Fig.~\ref{elias1data}.

The binarity of Elias~1 is important for the interpretation of the
X-ray flux observed from this object. HAeBe stars should lack the
convective zone necessary to drive a conventional dynamo. Novel dynamo
mechanisms have been developed to explain HAeBe activity, such as the
differential rotation dynamo model of Tout \& Pringle (1995). Also, it
has been suggested that deuterium burning in a shell outside a
radiative core powers a surface convective layer (Palla \& Stahler
1990).  Such models are of course not necessary if a low-mass
companion is in fact the source of the X-ray emission.  The spectral
type of A6 and luminosity of Elias~1 suggest, from comparison with 
tracks (Palla \& Stahler 1993), that its mass is of order
2--2.5$M_{\odot}$. The flux ratio of the system suggests a companion
mass of 1.5--2$M_{\odot}$, which lies close to the boundary between
fully radiative and partially convective stars. Therefore, it is
possible that the companion has a normal convection-generated magnetic
field and corona and is the source of the X-ray emission.

\subsection{HK~Ori}
\begin{figure*}[!t]
\vbox{%
\hbox{%
\psfig{figure=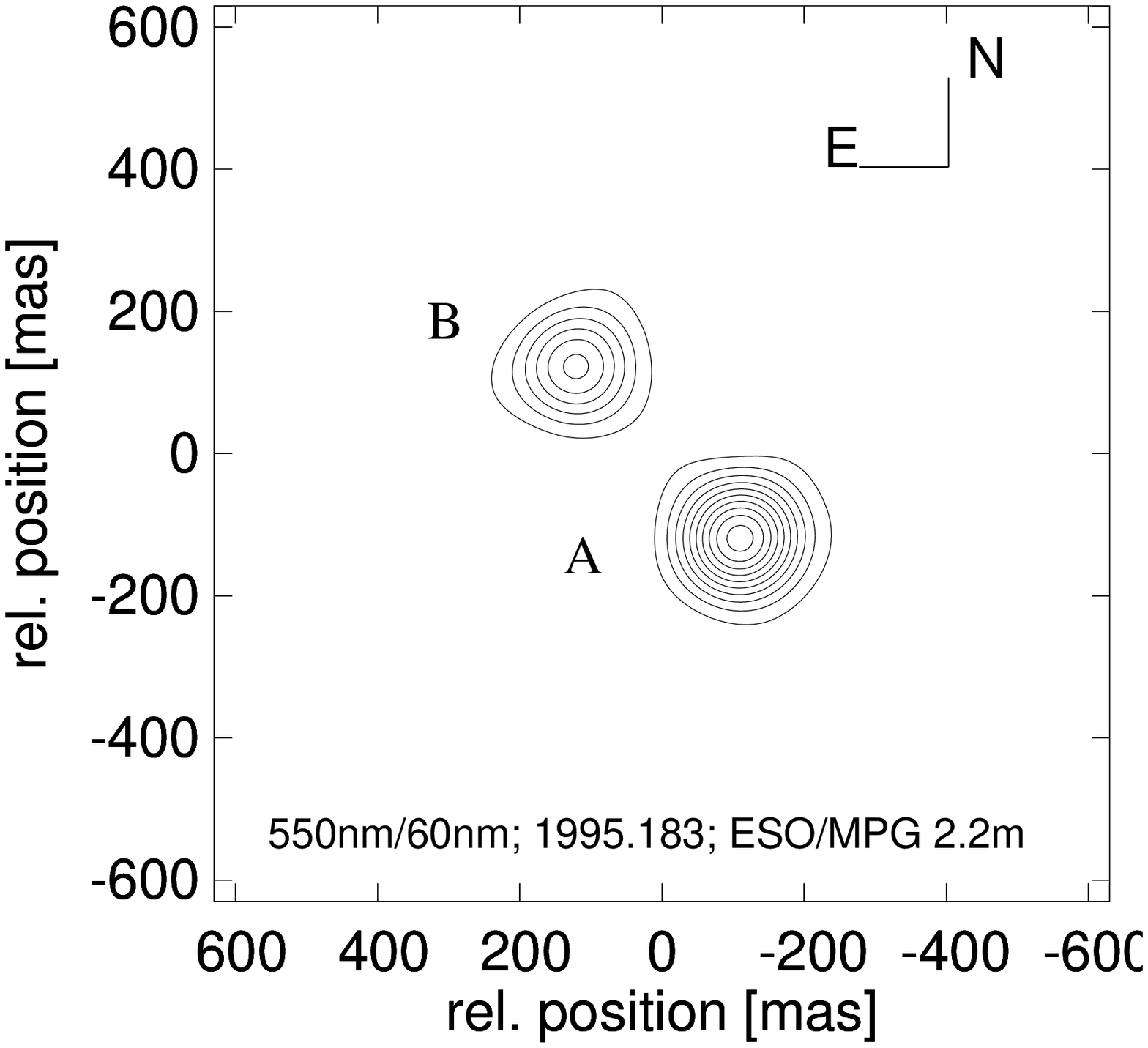, width=7.0truecm} 
\hspace{1.truecm}%
\epsfig{figure=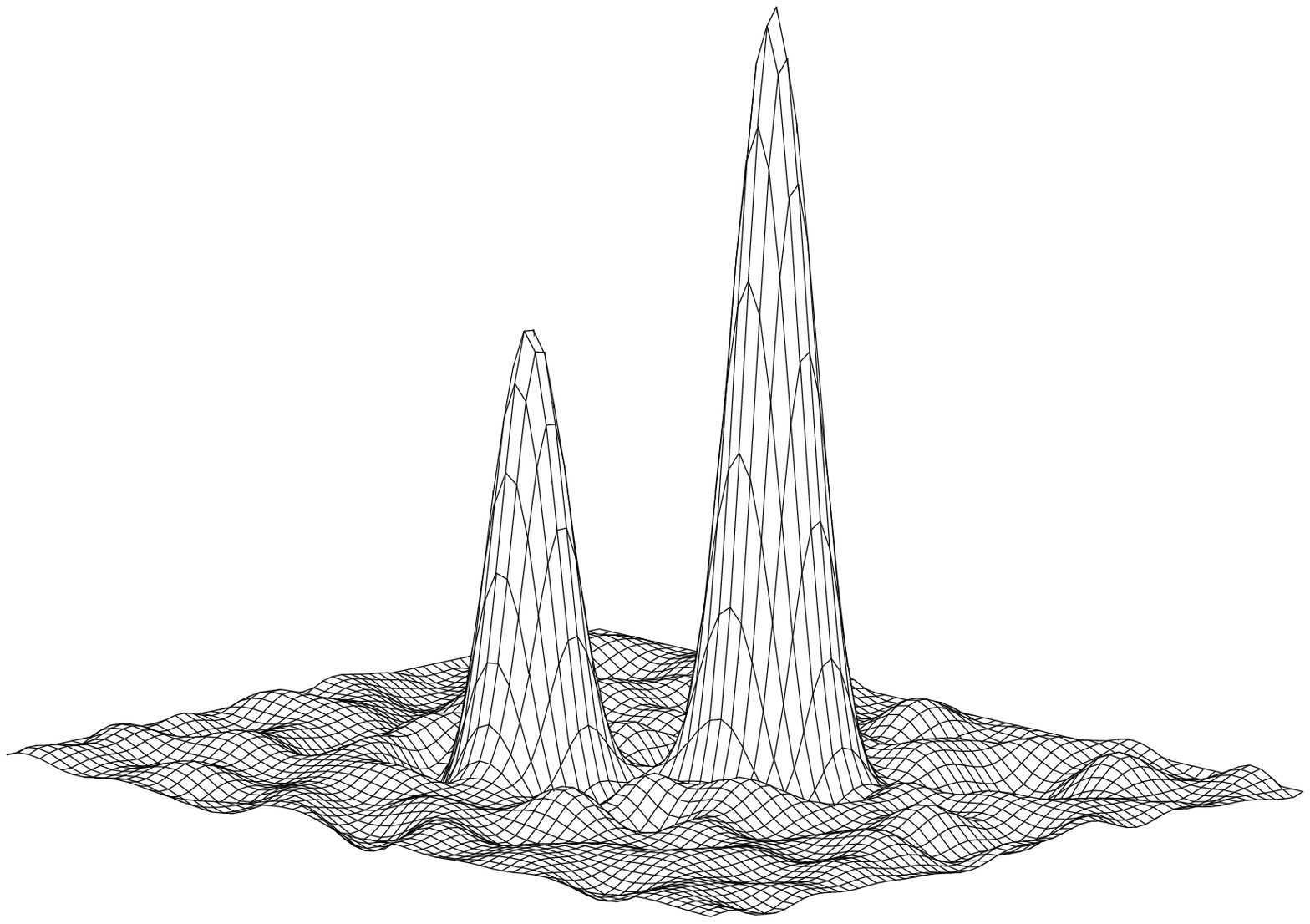, width=9.0truecm,bbllx=100,bblly=-50,bburx=651,bbury=325}
}
\hbox{%
\psfig{figure=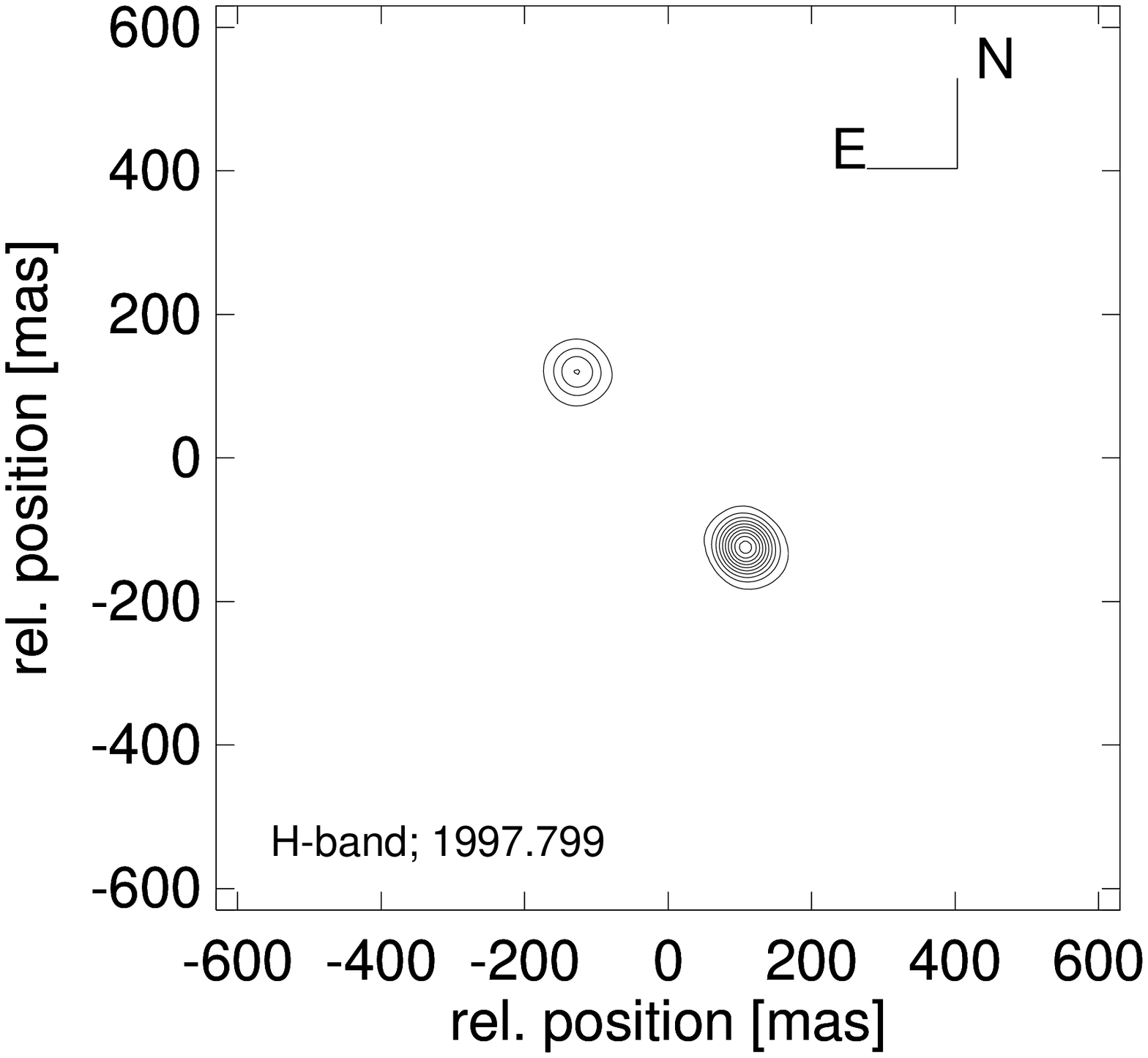, width=7.0truecm}
\hspace{1.truecm}%
\epsfig{figure=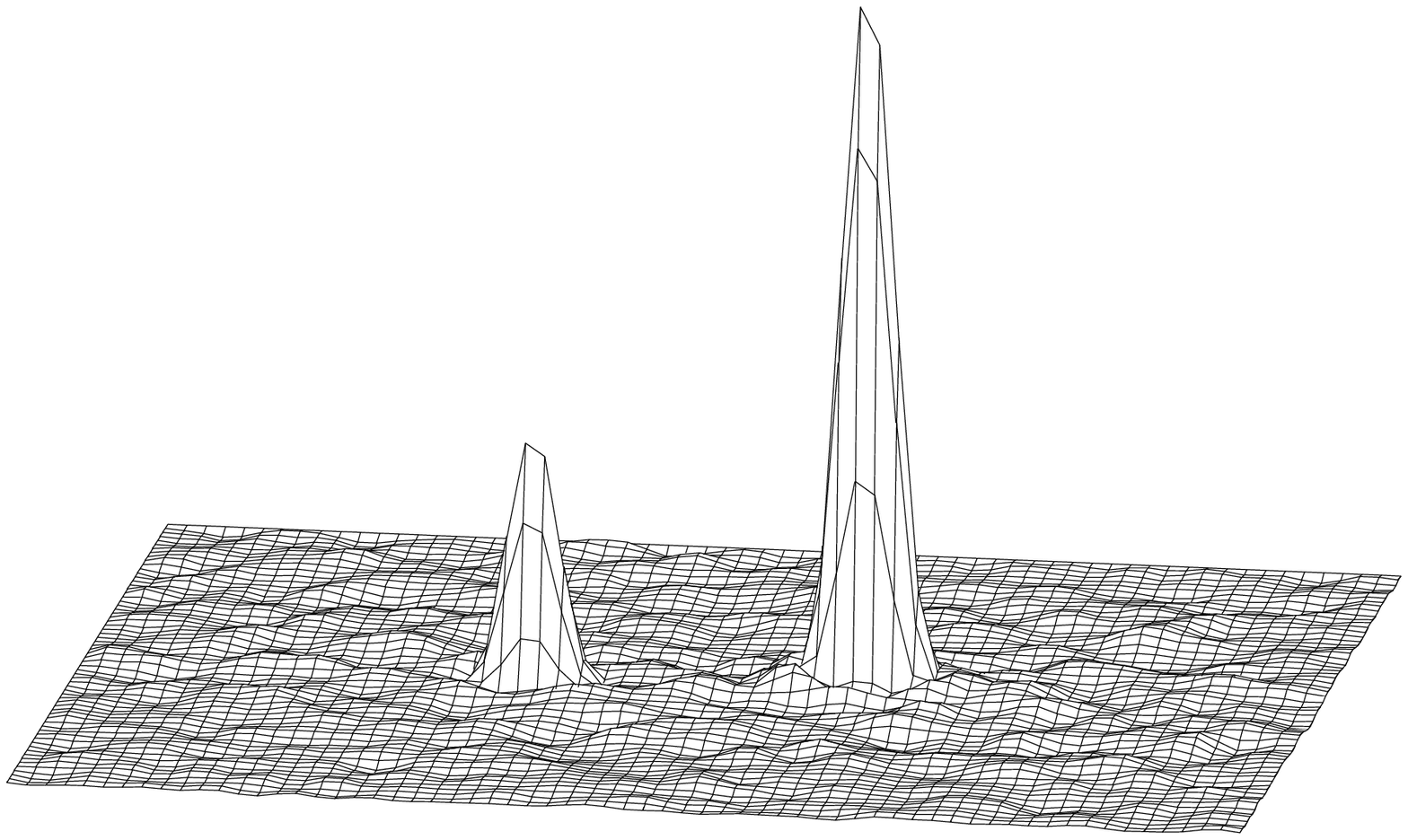, width=9.truecm,bbllx=100,bblly=-50,bburx=651,bbury=325}
}
\hbox{%
\psfig{figure=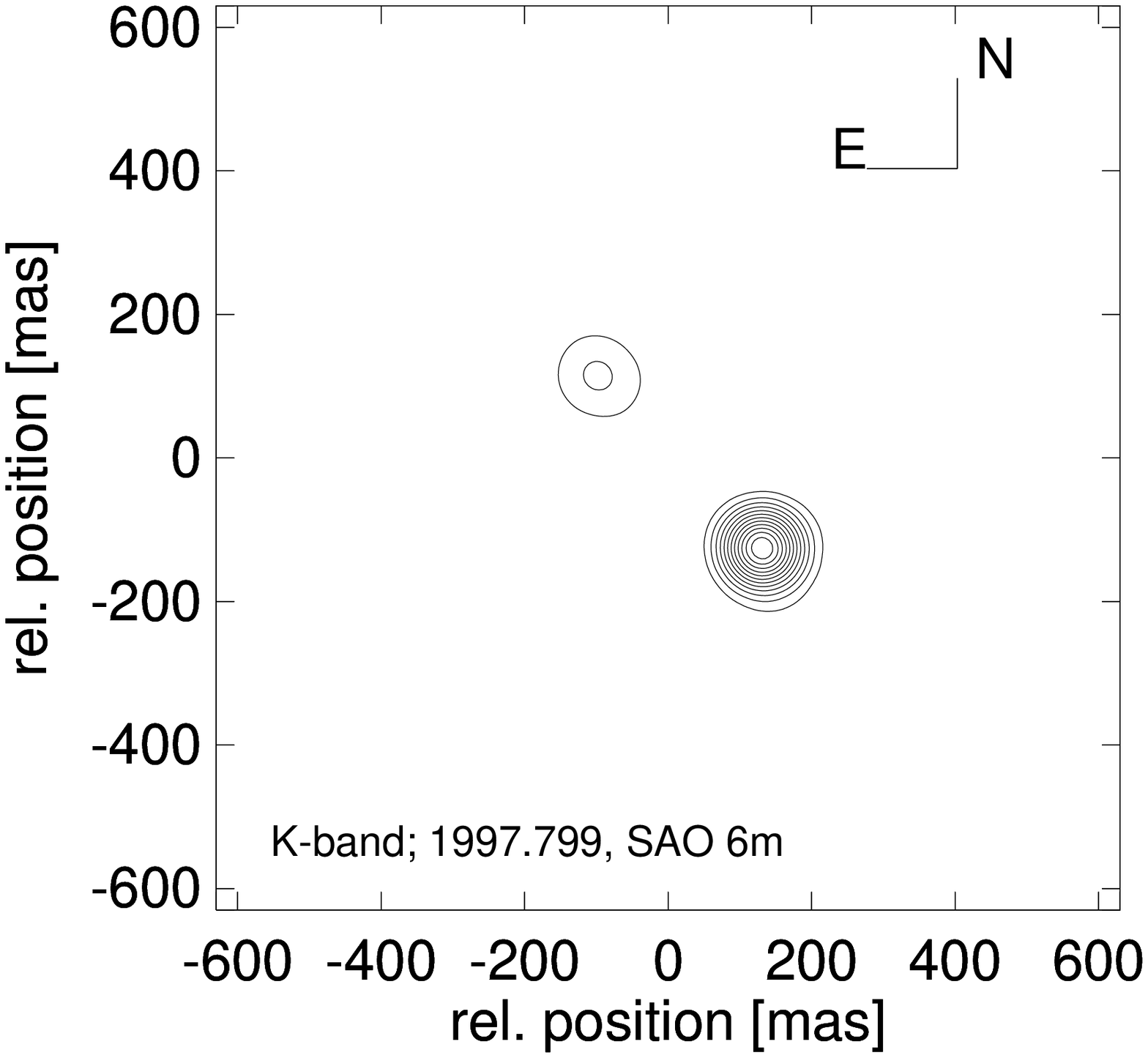, width=7.0truecm}
\hspace{1.truecm}%
\epsfig{figure=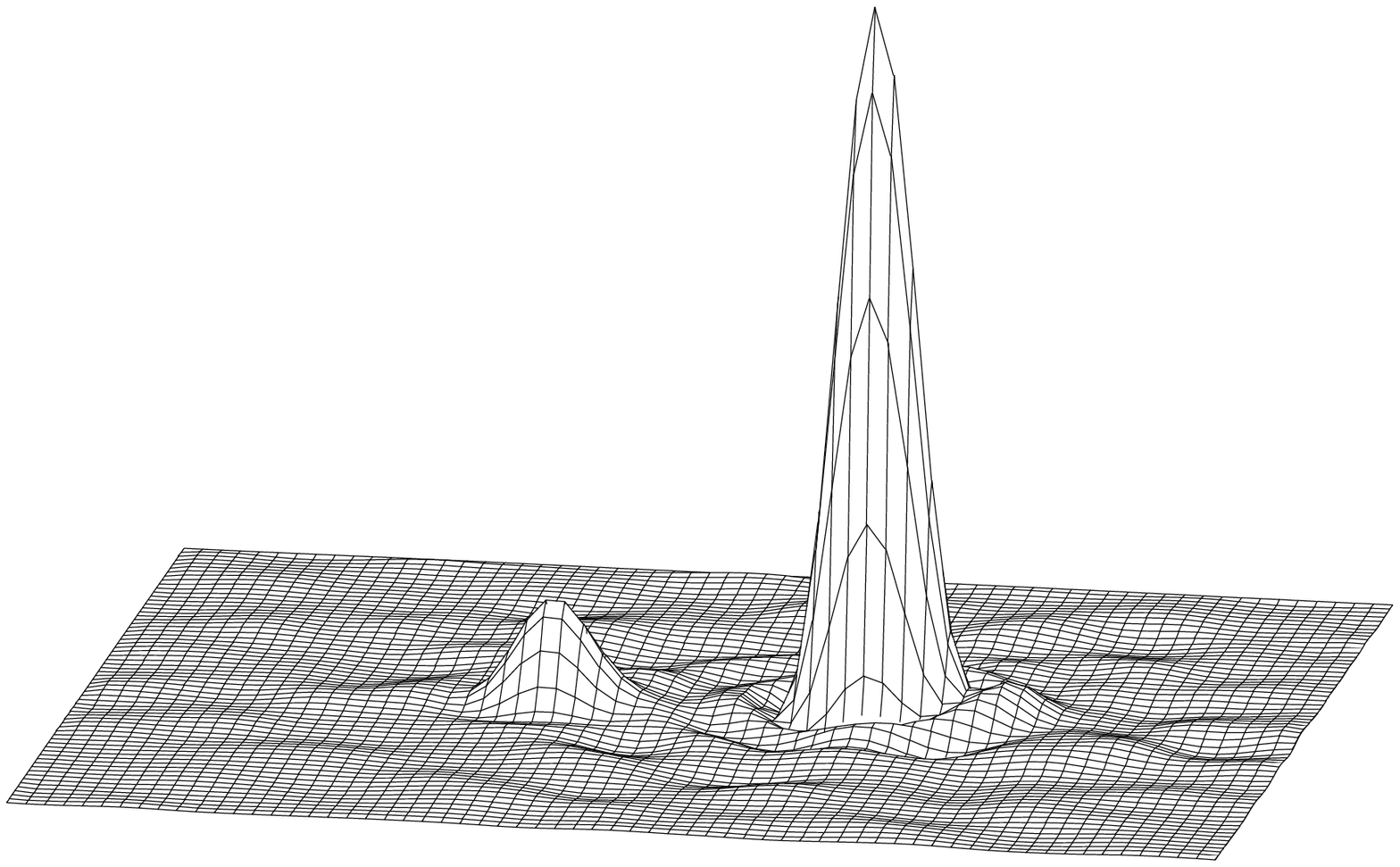, width=9.truecm,bbllx=100,bblly=-50,bburx=651,bbury=325}
}
}
{\caption[]{
Bispectrum speckle interferometry reconstructions of HK~Ori. Contour plots 
on the left show the relative positions of the sources and wire-frame plots on
the right illustrate the relative brightnesses. The components A and B  
are identified in the topmost contour plot.  }
\label{hkori3d}}
\end{figure*}

\subsubsection{The data}

\begin{figure}[t]
  \centering
   \psfig{figure=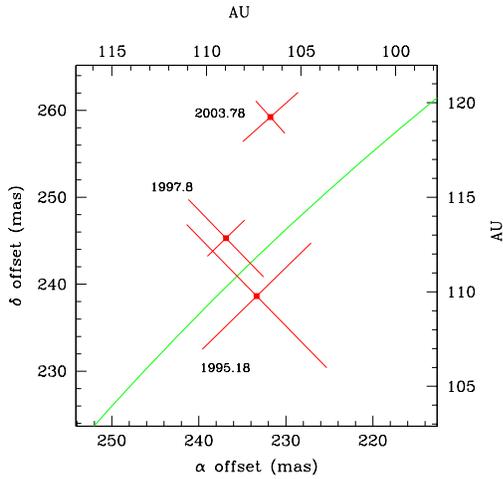,width=7.0truecm} 
   \caption{The position of the
HK Ori companion B relative to the primary. The separate optical
positions from 1995 and separate $H$ and $K$ positions for epoch 1997.8
have each been averaged to produce one position for each date.  The
solid line is a circle of radius 155~AU for an assumed distance of
460~pc centred on the primary.}
   \label{hkoridata}
\end{figure}

HK~Ori ($\alpha = 05^\mathrm{h} 31^\mathrm{m} 28\fs04$, $\delta =
+12\degr 09\arcmin 10\farcs3$, J2000) is an A5 type star with $A_V =
1.2$. It was found to be a binary by LRH97, who designated the
brighter NIR component {\bf to the southwest} A and the fainter {\bf
northeastern} component B.  LRH97 found a binary separation of
$(0.34\pm0.02)''$ and position angle $(41.7\pm0.5)^{\circ}$.  {\bf
Contour maps of the system in bands $V$, $H$ and $K$ are shown in
Fig.~\ref{hkori3d}, together with 3-d plots illustrating the
brightness ratio.  The track of component~B with respect to component~A 
is shown in Fig.~\ref{hkoridata}.}  Our data points trace a path
from southeast to northwest. The radial position of LRH97 is broadly
consistent with our measured positions, but the azimuthal position is
not in agreement with the position suggested for epoch 1992.12 from
our data.

Following LRH97, we can use our flux measurements to disentagle the
SEDs of the individual components of HK~Ori.  LRH97 used observations
at wavelengths of $I'$ (0.917\,\mic), $J$, $H$, and $K$, and had to
rely on flux ratios alone. We have data at four different wavelengths,
550~nm ($V$), 656~nm ($R$), as well as $H$ and $K$. The optical points
were not available to LRH97, and we are therefore able to separate the
SEDs of the components over a significantly wider wavelength
range. Furthermore, our data is photometrically calibrated (except for
the 2003 point which we do not use in this analysis). There remains
the problem that our measurements are not simultaneous. However,
independent measurements in October, 1997 show that the $V$-magnitude
was the same as during our earlier observations in 1995
($V$=11.7$\pm$0.1 derived from speckle observations in 1995,
$V$=11.6$\pm$0.1 in 1997, Mel'nikov private communication). This
indicates that the system was in a comparable state at both epochs and
that it is therefore valid to combine our data into one SED. 

In Fig.~\ref{hkori_sed} we present the separated SED (along with
various models described in Sect.~\ref{sec:hkori:models}).  SEDs for
the whole system were taken from Hillenbrand et al. (1992) and
Berrilli et al. (1992).  An upper limit at 1.3~mm was taken from
Henning et al. (1997). The error bars for $U$, $B$ and $V$ reflect the
measured variability as reported by Eiroa et al. (2002) in the
$V$-band. For $R$ through $K$ the measurement uncertainty was used
(typically of order 10\%), and for wavelengths longer than 2.2\,\mic,
an assumed 10\% uncertainty was adopted.

The SEDs of the individual components A and B are constructed from points
measured by us at $V$, $R$, $H$, and $K$, and points measured by LRH97 at
$I_J$, $J$, $H$ and $K$. Our points are calibrated as described in
Sect.~\ref{observations}. LRH97 obtained flux ratios but not absolute fluxes.
We therefore normalised their values against the total flux given by
Hillenbrand et al. (1992).  In the case of LRH97's $I_J$ point, the Cousins
system flux of the composite SED was converted to a supposed Johnson system $I$
flux using the information given in Bessell (1983). 

All fluxes were corrected for the effect of interstellar reddening.  The visual
extinction, $A_V=1.2$ was determined by Hillenbrand et al. (1992) from the $(B
- V)$ excess.  The extinction law given by Cardelli et al (1989) was used to
estimate the wavelength dependency of the extinction, with the assumption that
$R_V=3.1$.

The SED is clearly dominated by component A (which we designate the
primary) at longer wavelengths. The brightness ratio moves towards
unity in the very near-infrared ($I$ and $J$ bands).  LRH97 considered
one possible scenario in which the brightest infrared component, A,
might in fact be a low-mass IR companion, and the fainter component
the primary.  {\bf This conjecture was apparently confirmed by Baines
et al. (2004) who used a spectroastrometry technique to show that the
southwestern component dominates the emission at H$\alpha$, 
Their data also suggested that the northeastern component 
might be the brightest optical source in the system. 
However, our data points at 550~nm and 656~nm now exclude
this possibility, and show clearly that the brighter IR component {\bf
to the southwest} also dominates in the optical {\bf (see top panel of
Fig.~\ref{hkoridata}). Variability of the T Tauri source 
is a possible but not convincing explanation for this discrepency, since 
the photometrically calibrated flux of component A matches the 
composite SED published in Hillenbrand et al. (1992), whereas the 
flux of component B does not. }
\begin{figure*}[!ht]
   \centering
   \psfig{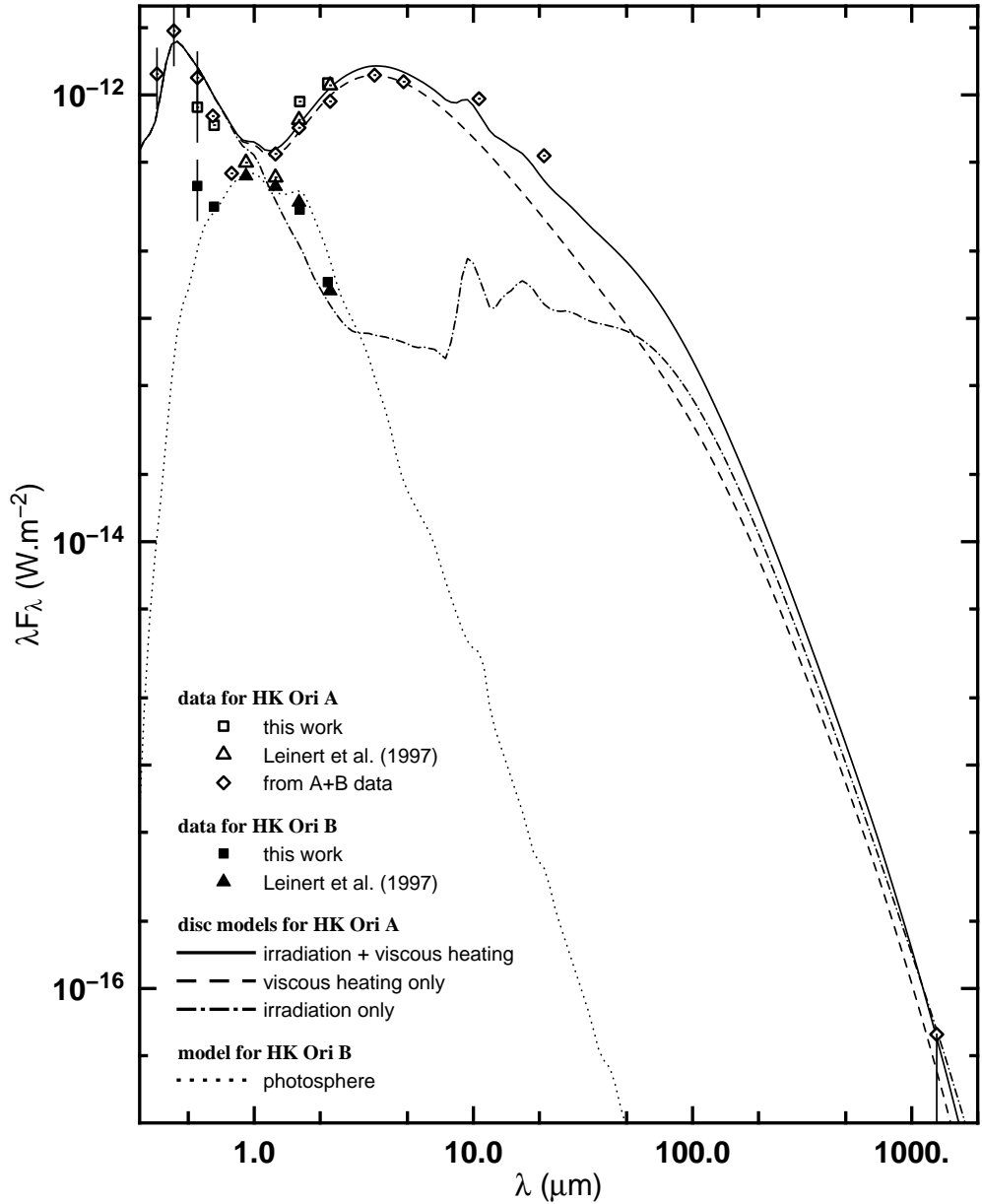}
   \caption{The SED of HK~Ori's components.  \strong{Data:} Open
   symbols stand for the primary and filled ones for the
   secondary. Separate photometric measurements for HK~Ori~A and
   HK~Ori~B from this work are marked with squares. Flux ratios
   measured by LRH97 and normalised to separate fluxes by reference to
   the composite photometry of Hillenbrand et al. (1992) are marked
   with triangles.  For other points (diamonds), we have subtracted
   the model fit to HK~Ori~B from the total system flux as measured by
   Hillenbrand et al. (1992) to obtain supposed values for the flux of
   HK~Ori~A.  We also show the 1.3~mm upper limit obtained by Henning
   et al. (1998), also with the flux of the HK~Ori~B model
   subtracted. In $U$, $B$, and $V$, the uncertainties result from the
   reported variability. At NIR wavelengths, the measurement
   uncertainties are of order 10\% and are not plotted for the sake of
   clarity.  \strong{Model for HK~Ori~A:} A two-layer disk model by
   Lachaume et al.  (2003), including irradiation by the central star
   and heating by viscosity fits the data very well at all wavelengths
   (solid line).  For comparison's sake similar models, in which
   irradiation or viscous heating is turned off, are displayed (dashed
   line is without irradiation and dash-dotted line is without viscous
   heating).  Model parameters are shown in Table~\ref{fittable}.
   \strong{Model for HK~Ori~B:} The SED of the secondary does not hint
   at an infrared excess, so a photosphere model is fit to the data
   (dotted lines).  It should be noted that excess emission from
   distant reprocessing material cannot be ruled out.}
   \label{hkori_sed}
\end{figure*}

The infrared colours of the two components are very different. The
brightest component has $H - K = 2.12$, whereas component B has $H - K =
1.10$. The implication is that component A possesses significantly
more circumstellar material than component B, which resembles a naked
photosphere of around 4000~K. The positions of the two objects are
shown in a colour-colour diagram in Fig.~\ref{colcol}. Comparing this
figure to colour-colour diagrams of young stars, for example Fig.~15
of Hillenbrand et al. (1992), it is clear that the primary (component
A, to the southwest) occupies a typical position for an intermediate
mass young star, whilst the companion (component B, to the northeast)
occupies a position typical for a T Tauri star.

We have fitted a 4000~K photosphere, corresponding to a mid-K spectral
type, to the SED of component B.  This fit 
represents the SED of component~B very well, with the possible 
exception of the $B$-band point. Where no resolved points were
available, we have subtracted this fit to determine points for A. In
practice, this model subtraction makes little difference to most of
the fluxes measured for the composite system, and this method is
nearly equivalent to assuming that component A contributes essentially
all the flux at $U$, $B$ and in the mid-infrared. 

\subsubsection{Modelling strategy}
\label{sec:hkori:models}

Component~A is responsible for most of the IR-excess of the system and
presents a ``double-bumped'' SED.  The first bump peaks around 400\,nm and
is produced by the central A-type star.  The second one, peaking around 
3~\mic,  is a signature of the presence of circumstellar matter and
is usually interpreted as emerging from an accretion disk (e.g. Hillenbrand
et al., 1992).

Let us emphasise that the NIR- and MIR-excesses are unusually strong
for a Herbig star, with a flux intensity $\lambda F_\lambda$ peaking
as high as the stellar photosphere (see Fig.~\ref{hkori_sed}) while it
is one order of magnitude smaller in most HAeBes (see Fig.~1 in Natta
et al. 2001). Making a simple three blackbody fit, we estimate that
the circumstellar flux accounts for fully two thirds of the observed
flux from component~A.  {\bf The low visual extinction ($A_V = 1.2$)
hints that the star is not significantly shadowed by circumstellar
material, so that the circumstellar matter actually emits twice as
much flux as the star in the observer's direction. Higher values of
$A_V$ for the photosphere can be considered if the extinction law is
taken to be non-standard ($R_V \sim 5$ leads to $A_V \sim 3$ for the
measured $E(B - V)$), but we feel this is unlikely since the the
derived extinction is in agreement with the extinction from
interstellar matter measured for nearby Herbig-Haro objects (Goodrich
et al. 1993). } Though the reprocessing of stellar light is widely
accepted to be the main source of energy in typical HAeBes (Chiang \&
Goldreich 1997; Natta et al. 2001; Dullemond et al. 2001), it seems
unable to account for so high a IR-excess: an optically thin envelope
would reprocess only a fraction $\tau \ll 1$ of the stellar light, and
an optically thick disk $\approx h/r \sim 0.05\mbox{--}0.5$, where $h$
is the thickness of the disk, and $r$ the distance to the star.  So,
we follow Hillenbrand et al. (1992) in their assumption that viscous
heating (i.e.  accretion luminosity) is responsible for most of the
observed flux.  We use a generalised version of the Chiang \&
Goldreich (1997) model by Lachaume et al. (2003), that includes
viscous dissipation in addition to stellar irradiation, to fit the
SED. Our approach has the advantage of not excluding \emph{a priori}
the hypothesis of an irradiated disk.
\begin{figure}[!ht]
   \centering
   \psfig{figure=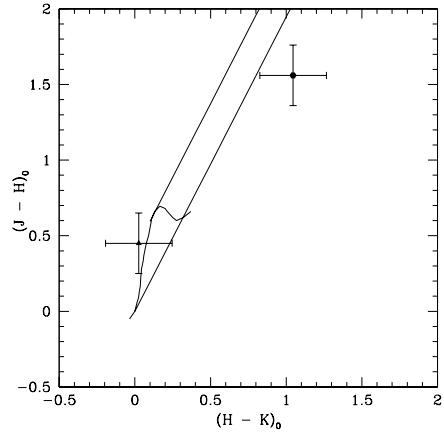,width=6.0truecm} 
   \caption{(H - K) vs (J - H) colour-colour diagram showing the positions of
   the primary and secondary in the HK Ori~system. Component A is shown with a
   circular point and component B with a triangle. The colours have been
   corrected for the effects of interstellar reddening. The main sequence (from
   Bessell \& Brett, 1988) and the direction of interstellar reddening are also
   shown.}
   \label{colcol}
\end{figure}

The fitted parameters are summarised in Table~\ref{fittable}, and the SED data
and fit are displayed in Fig.~\ref{hkori_sed}. The large IR excess at 
$ \lambda \gtrsim 2$~\mic\ cannot
be explained with irradiation only (dash-dotted line in
Fig.~\ref{hkori_sed}) and requires a high accretion rate $\Mdot =
2.5\,10^{-6}\,\Msunperyr$ (solid line).  The depression in the SED around
1~\mic\ suggests that the inner, hot regions of the disk should be depleted.
which our model backs:  the inner rim of the disk is located at 
$31\,\Rsun = 20\,R_{*} = 0.15\,\AU$ from the star. 

On one hand, the accretion rate and outer disk truncation radius
are in good agreement with those derived by Hillenbrand et al. (1992).
This could be expected, since these parameters are sensitive to the
mid- and far-IR SED that was not affected by our resolving the binary.
On the other hand, the visible and NIR SED for component A
significantly differs from that of the total system, so that our
estimates for the stellar photosphere (derived from the visible
photometry) and the inner truncation (sensitive to the depression at
1\,\mic) are quite different from their values.

\subsubsection{Disk properties}

The model was computed with the viscosity coefficient $\alpha = 2 \times
10^{-2}$ (see Shakura \& Sunyaev, 1973), but choosing any $\alpha$ above
$10^{-3}$ does not greatly affect the quality of the fit.  The upper limit at
1.3~mm constrains the amount of material at larger radii, and the mass of the
disk is limited to about $0.06\,\Msun$.  Combined with the high accretion rate,
this estimate leads to a lifetime of the disk between a few $10^{4}$ and a few
$10^{5}$ years, i.e. less than the typical age of HAeBe stars.  On one hand,
the lifetime may be extended by replenishing of the disk with material from
some wider reservoir.  The scenario of a circumbinary disk is unlikely, since
material with high angular momentum should accrete on to the secondary instead
of the primary, at least in a coplanar system.  This favors the hypothesis of
mass infall from a spherically symmetric envelope. This envelope should however
comprise at least a few solar masses, implying a significant reddening, which
is not observed.  On the other hand, HK~Ori could be in a transitory
high-accretion state of an otherwise ``quiescent'', passive accretion disk.

More concerning is the presence of a large central gap
($20\,\Rstar$) in an intensively accreting disk.  As pointed out by
Hartmann et al.  (1993), a strongly accreting disk should not present
such a gap: the gap is either understood as a drop in optical
thickness beyond the dust sublimation radius, or as a cleared region
due to stellar magnetic fields which drive the matter into accretion
streams or into an outflow.  However, the large amount of material in
this disk would ensure that the gas remains optically thick, whilst
the field strength required to truncate the disk at 20$R_{*}$ would be
of order tens of kilogauss. Though these arguments plead for a
moderately accreting disk, we were not able to find a mechanism
other than accretion that could account for the very large
IR-excess. In particular, the reprocessing of the stellar light leads
to IR fluxes of up to a few tens of percent of that of the star, as previous
simulations by Chiang \& Goldreich (1997) and Dullemond et al. (2001)
suggest.

\begin{table}
   \caption{Fitted parameters for HK~Ori. A disk mass of $0.06\,\Msun$ 
   is derived from these parameters. The parameters obtained by Hillenbrand
   et al. (1992) are given in italic when available.} 
   \label{fittable}
   \centering
   \begin{tabular}{lll}
      \hline \hline
      \multicolumn{3}{c}{Component A}\\
         stellar radius (\Rsun)       &  1.55 & \textit{1.7}$^d$\\
         stellar mass (\Msun)         & 2.0          & \textit{2.0}$^d$\\
         stellar temperature (K)      &  8500 & \textit{9700}$^d$\\
      \multicolumn{3}{c}{Component B}\\
         stellar radius (\Rsun)       & 4.1 \\
         stellar mass$^a$ (\Msun)     & 1.0 \\
         stellar temperature (K)      & 4000\\
      \multicolumn{3}{c}{Disk around A}\\
         disk inner radius (\Rsun)    & 31            & \textit{17}\\
         disk outer radius (\AU)      & 30            & \textit{29}\\
         accretion rate (\Msunperyr)  & \sci{2.5}{-6} & $\mathit{\sci{1.7}{-6}}$\\
         $\alpha${}                   & 0.02          & ---\\
         disk inclination$^a$ (deg)   & 0             & \textit{0}\\
         distance$^b$ (pc)            & 460           & \textit{460}\\
      \hline
      \multicolumn{2}{l}{$^a$ kept constant}\\
      \multicolumn{2}{l}{$^b$ assumed}\\
      \multicolumn{2}{l}{$^c$ derived from simulation}\\
      \multicolumn{2}{l}{$^d$ both components}\\
      \hline\hline
   \end{tabular}
\end{table}

\subsection{V380 Ori}

V380~Ori ($\alpha$= 05$^h$~36$^m$~25.43$^s$,
$\delta$=-06$^{\circ}$~42$'$~57.7$''$, J2000) is a B9 star in the
Orion complex at a distance of 460~pc. It was classified as a group~I
source by Hillenbrand et al. (1992), meaning that its SED can be
fitted by a geometrically thin, optically thick accretion disk.

The system was resolved as a $\sim$150~mas binary by LRH97.  
They attempted to divide the SED of the system into the
contributions of the two components. As with HK~Ori, they considered
two models; one in which the brightest source in the IR continues to
dominate in the optical, and one in which the brighter IR source is an
``IR companion'' which is less bright in the optical. Millan-Gabet et
al. (2001) used long-baseline interferometry to resolve
V380~Ori~A. Their uniform ring model fit had an inner radius of
2.5~mas, corresponding to just over 1~AU. 

In Fig.~\ref{v380data} we show a contour plot of the V380~Ori
system, and a plot of the position we obtained for the secondary in
2003, together with the position obtained by LRH97.

\begin{figure}[ht]
\vbox{ 
\begin{center}
\psfig{figure=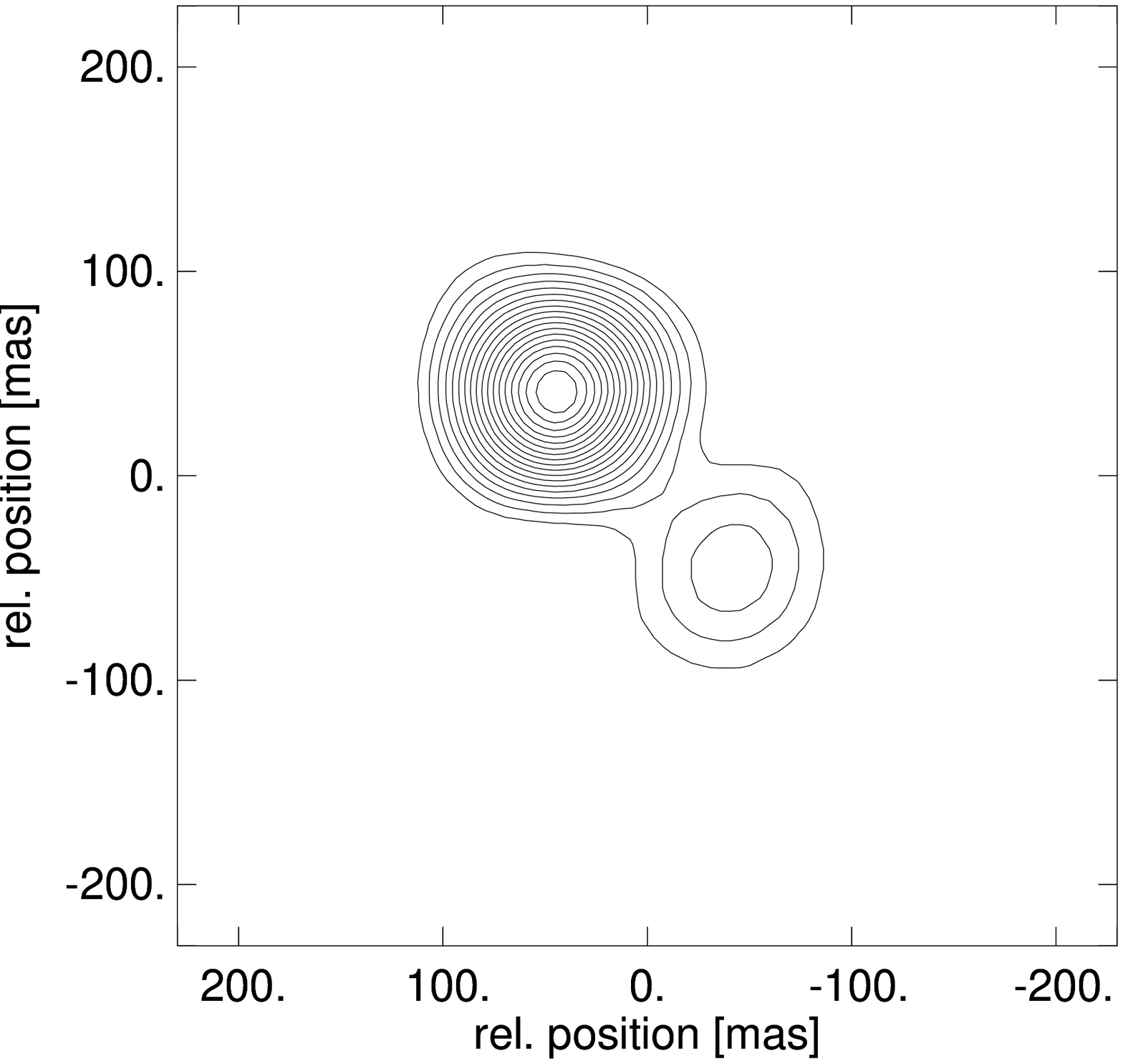,width=6.0truecm}
\psfig{figure=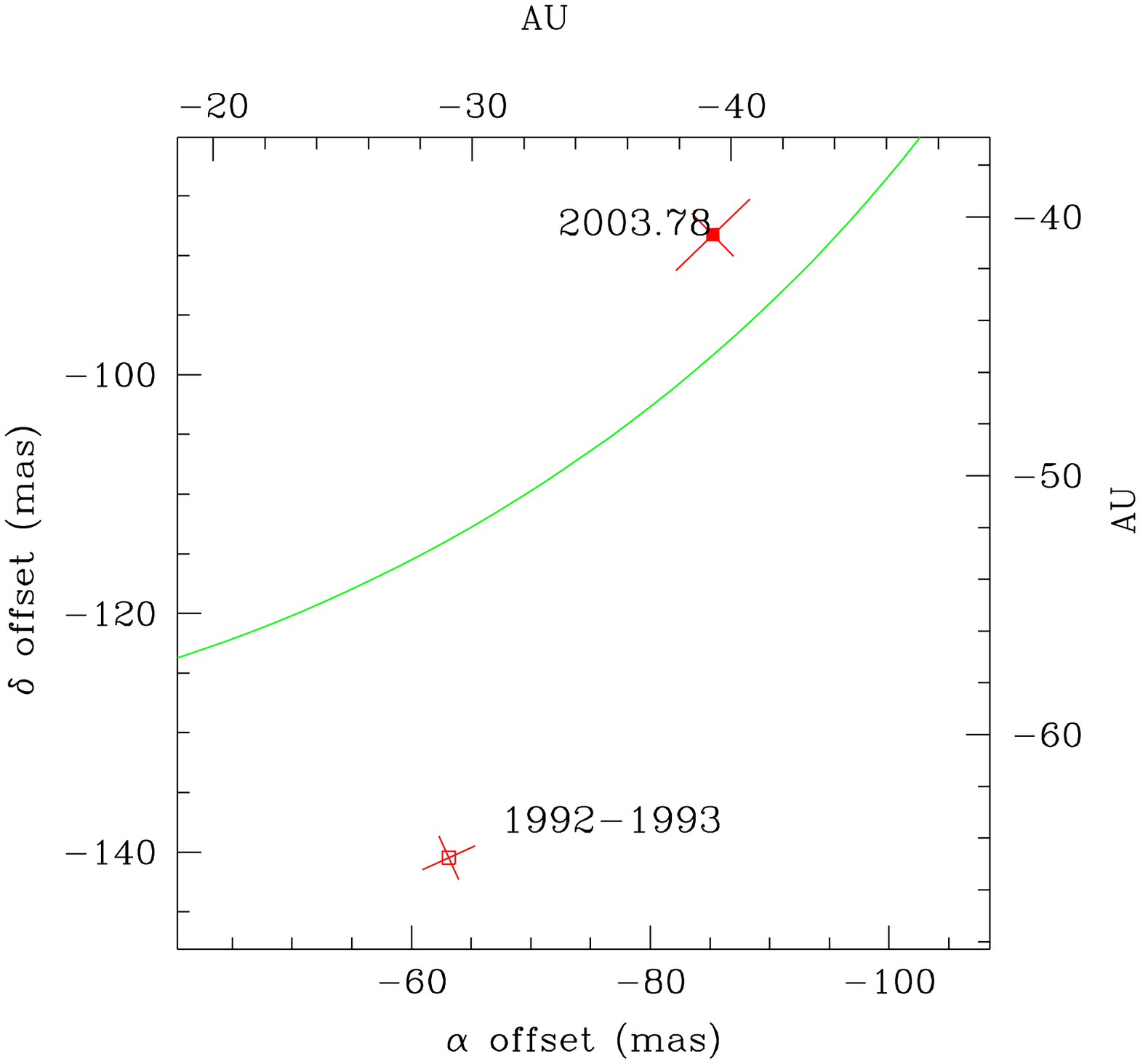,width=7.0truecm} 
\end{center}
} 
{\caption{Top: Contour plot of the V380~Ori system. Bottom: Position of the
   V380~Ori companion relative to the primary. The solid line is a
   circle of radius 60~AU centred on the primary. The data point of
   LRH97 is shown with an open diamond. This position
   was obtained by LRH97  from data taken over a period of 3
   years. Our data point is marked with a filled symbol.}
           \label{v380data}}
\end{figure}

Based on the data presented, it would appear that there has been
orbital movement of the V380~Ori system.  The relative position of the
secondary has changed by about 26~AU. This apparent shift has been
largely in a tangential sense, as can be seen by comparison with the
circle segment shown in the figure. 

The flux ratio at K$'$ is measured by us to be 0.266$\pm$0.008, whereas LRH97 
obtained 0.35$\pm$0.08. These values are almost compatible with each other to within $1\sigma$.
Unfortunately, unlike in the case of HK~Ori, we have no optical 
observation and so cannot resolve the ambiguity considered by these authors 
in the SEDs.

\section{Conclusions}

We have used bispectrum speckle interferometry to resolve for the
first time the binary companions to the HAeBe stars \lkha\ and
Elias~1. In the case of \lkha, the new object lies at a separation of
approximately 60~mas, or 36~AU from the brighter component. Our
observations comprise seven data points taken at intervals of
approximately one year over a seven year time period. Motion of the
secondary with respect to the primary is seen and the path is
suggestive of an arc, although a straight line cannot be excluded.
Because the companion may well have an important effect on the
interesting and well-studied circumstellar environment of \lkha, we
have fitted several possible orbits to these points and shown that the
plane of the orbit cannot be viewed {\bf exactly} edge-on {\bf with an
exactly east-west orientation}. This may imply that the binary is not coplanar
with the inferred disk powering the outflow. The flux ratio of the
Elias~1 system indicates that the new companion has a mass of around
2$M_{\odot}$ and therefore may be a convective star. This opens the
possibility that the X-ray emission of Elias~1 originates from the
companion and is powered by conventional mechanisms. The SED of HK~Ori
was separated into the contributions of the primary and secondary
components. The IR excess of HK~Ori~A was found to contribute around
two thirds of the total emission from this component, suggesting that
{\it accretion power} contributes significantly to the flux, in
contrast to most HAeBe stars whose SEDs can often be well fitted with
passive disks.  We fitted an accretion disk model to the HK Ori A SED
and determined the accretion rate and disk mass. From these
parameters, the inferred disk lifetime is found to be quite short
compared to the suspected age of HK~Ori. It is possible that the disk
could be replenished from a wider reservoir of material, but there are
problems with such a picture; a spherical envelope should lead to
extinction which is not observed, whilst in a coplanar system,
accretion from a disk should proceed preferentially onto the
secondary. Another explanation could be that the HK~Ori disk is seen
in an outburst phase. Such a phase need not affect the outer disk, so
that the accretion rate may not be uniform throughout.  Our relative
positions for this system trace a path over the eight years of
observations which may be the segment of an orbit. {Future
near-simultaneous multi-wavelength high resolution observations will
allow a more detailed picture of the disk}.  
High-resolution observations in the mid-infrared could determine whether
or not component B has a passive disk.
Further observations of
all four stars will be necessary to build up reliable orbit
determinations. \lkha\ in particular should be a fascinating subject
for future study from this point of view, since the close binary
companion may be responsible for much of the complexity in the
circumstellar environment of this system.

\begin{acknowledgements}

We would like to thank the anonymous referee for 
constructive comments which 
have helped us to improve the manuscript.
				
\end{acknowledgements}

\end{document}